%% file: arXiv.tex
\documentclass[aps,prb,superscriptaddress,twocolumn,10pt,longbibliography]{revtex4-2}
\usepackage{amsmath, amssymb}
\usepackage{braket}
\usepackage{color}
\usepackage{graphicx}
\usepackage{ulem}
\usepackage{siunitx}

\begin{document}
\date{\today}
\title{Tunnel magnetoresistance effect with a Cr-doped $\mathrm{RuO_{2}}(110)$ altermagnet}
\author{Katsuhiro Tanaka}
\affiliation{Department of Materials Design and Engineering, University of Toyama, Gofuku, Toyama-shi, Toyama 930-8555, Japan}
\author{Takuya Nomoto}
\affiliation{Department of Physics, Tokyo Metropolitan University, Hachioji, Tokyo 192-0397, Japan}
\author{Ryotaro Arita}
\affiliation{Department of Physics, University of Tokyo, Hongo, Bunkyo-ku, Tokyo 113-0033, Japan}
\affiliation{Center for Emergent Matter Science, RIKEN, Wako, Saitama 351-0198, Japan}
\begin{abstract}
Antiferromagnets can have a finite spin-polarization in the momentum space when their magnetic structure breaks the macroscopic time-reversal symmetry.
This spin-polarization can produce a spin-polarized electric current even in antiferromagnets with vanishingly small net magnetizaton, 
which supports the antiferromagentic tunnel magnetoresistance (TMR) effect.
In this paper, using first-principles calculations, we study the TMR effect with a doped altermagnet $\mathrm{Ru}_{1-x}\mathrm{Cr}_{x}\mathrm{O}_{2}$ with $(110)$ orientation,
whose collinear antiferromagnetic structure breaks the time-reversal symmetry macroscopically.
The momentum-dependent spin-polarization combined with the $(110)$ crystal orientation makes the electric current spin-polarized through bulk $\mathrm{Ru}_{1-x}\mathrm{Cr}_{x}\mathrm{O}_{2}(110)$.
We further calculate the TMR effect in the $\mathrm{Ru}_{1-x}\mathrm{Cr}_{x}\mathrm{O}_{2}(110)/\mathrm{TiO_{2}}(110)/\mathrm{Ru}_{1-x}\mathrm{Cr}_{x}\mathrm{O}_{2}(110)$ tunnel junction and show that a finite TMR effect emerges.
Based on the analysis of the tunneling transport,
the TMR effect is attributed to the spin polarized tunneling transport with momentum dependence and the interfacial magnetic structures, as well as the spin-polarized electric current in a bulk form of $\mathrm{Ru}_{1-x}\mathrm{Cr}_{x}\mathrm{O}_{2}(110)$.
\end{abstract}
\maketitle
\section{Introduction}
\label{sec:introduction}
When the electric current flows through the ferromagnetic metals, the current can be spin-polarized.
The tunnel magnetoresistance (TMR) effect utilizes such spin-polarized current~\cite{Julliere1975_PhysLettA_54A_225}.
Owing to the spin-polarized current tunnels through the magnetic tunnel junctions (MTJs),
which are constructed by two magnetic electrode layers sandwiching an insulating spacer,
the difference of the resistance can be generated between the parallel and antiparallel alignments.
The TMR effect has been observed generally in the MTJs based on ferromagnetic electrodes~\cite{Miyazaki1995_JMagnMagnMater_139_L231,Moodera1995_PhysRevLett_74_3273,Butler2001_PhysRevB_63_054416,Mathon2001_PhysRevB_63_220403,Parkin2004_NatMater_3_862,Yuasa2007_JPhysD_40_R337}.
\par
Meanwhile, recent studies have demonstrated that antiferromagnets can also generate a spin-polarized current thanks to its spin-polarization in the momentum space, 
despite the absence of the net magnetization, when their magnetic orders break the macroscopic time-reversal symmetry~\cite{Zelezny2017_PhysRevLett_119_187204,Zhang2018_NewJPhys_20_073028,Naka2019_NatCommun_10_4305,Gonzalez-Hernandez2021_PhysRevLett_126_127701,Watanabe2024_PhysRevB_109_094438}.
When such a spin-polarized current from antiferromagnets flows through insulating barrier,
the tunneling current can be also spin-polarized, resulting in the TMR effect, in analogy with the ferromagnetic MTJs.
Actually, the emergence of the tunnel magnetoresistance (TMR) effect with antiferromagnets whose magnetic configurations break the time-reversal symmetry has been discussed experimentally and theoretically~\cite{Zelezny2017_PhysRevLett_119_187204,Shao2021_NatCommun_12_7061,Smejkal2022_PhysRevX_12_011028,Dong2022_PhysRevLett_128_197201,Chen2023_Nature_613_490,Qin2023_Nature_613_485,Shao2023_PhysRevLett_130_216702,Cui2023_PhysRevB_108_024410,Jiang2023_PhysRevB_108_174439,Chi2024_PhysRevApplied_21_034038,Samanta2024_PhysRevB_109_174407,Zhu2024_ChinPhysLett_41_047502,Chou2024_NatCommun_15_7840,Zhang2024_PhysRevB_110_024428,Tanaka2024_PhysRevB_110_064433,Chou2024_NatCommun_15_7840,Gurung2024_NatCommun_15_10242,Luo2025_PhysRevB_111_144417,Wang2024_ApplPhysLett_125_202404,Luo2025_PhysRevB_111_144417,Yang2025_Newton_1_100142,Liu2025_AdvSci_12_e02985,Noh2025_PhysRevLett_134_246703,Zhu2025_ApplPhysLett_127_082401,Sun2025_PhysRevB_112_094411,Yang2025_PhysRevB_112_205202,Luo2026_AdvFunctMater_36_e28671,Tanaka2026_PhysRevMater_10_044405,Mao2026_PhysRevB_113_174409,Guo2026_ACSNano_20_18900,Shao2024_npjSpintronics_2_13,Tanaka2025_JPhysCondensMatter_37_183003,Kang2025_arXiv_2509.03026,Elekhtiar2026_arXiv_2605.25369}.
\par
The spin polarization in the momentum space and resultant spin-polarized current can be highly anisotropic depending on the crystal orientation, especially in antiferromagnets.
One of the representatives is the rutile $\mathrm{RuO_{2}}$.
The magnetism of $\mathrm{RuO_{2}}$ is under debate even now as several studies have suggested a nonmagnetic state ~\cite{Hiraishi2024_PhysRevLett_132_166702,Smolyanyuk2024_PhysRevB_109_134424,Kessler2024_npjSpintronics_2_50,Liu2024_PhysRevLett_133_176401,Osumi2026_PhysRevB_113_085116},
but once the collinear antiferromagnetic order appears,
its magnetic structure breaks the time-reversal symmetry macroscopically, namely, the altermagnetic state is realized~\cite{Berlijn2017_PhysRevLett_118_077201,Zhu2019_PhysRevLett_122_017202,Ahn2019_PhysRevB_99_184432,Smejkal2020_SciAdv_6_eaaz8809,Smejkal2022_PhysRevX_12_031042,Smejkal2022_PhysRevX_12_040501,Fedchenko2024_SciAdv_10_eadj4883}.
Then, the spin current generated by $\mathrm{RuO_{2}}$ shows anisotropic characters depending on the crystal orientation~\cite{Gonzalez-Hernandez2021_PhysRevLett_126_127701,Bai2022_PhysRevLett_128_197202,Karube2022_PhysRevLett_129_137201,Bai2023_PhysRevLett_130_216701}.
This anisotropy of the spin current can also affect the TMR effect; 
it has been proposed that the (110)-stacking can be more effective in producing a larger TMR effect~\cite{Jiang2023_PhysRevB_108_174439} compared with the (001)-stacking case~\cite{Shao2021_NatCommun_12_7061}.
In the (001)-stacking case, the total transport spin-polarization cancels out between up and down spin channels~\cite{Shao2021_NatCommun_12_7061}.
By contrast, in the (110)-stacking case, the spin polarization of the electric current does not cancel out owing to the crystal orientation,
and the spin-polarized current can also flow in the bulk form.
\par
Also in a derivative of the rutile $\mathrm{RuO_{2}}$ generated by chemical substitution of Ru by Cr~\cite{Wang2023_NatCommun_14_8240,Tanaka2024_PhysRevB_110_064433,Smolyanyuk2025_PhysRevB_111_064406},
the TMR effect has been investigated.
In Ref.~\cite{Tanaka2024_PhysRevB_110_064433}, it has been shown that Cr-doped $\mathrm{RuO_{2}}$ can work as the altermagnetic electrode of MTJs;
first-principles calculations have revealed that a finite TMR effect can appear in $\mathrm{Ru}_{1-x}\mathrm{Cr}_{x}\mathrm{O}_{2}(001)/\mathrm{TiO_{2}}(001)/\mathrm{Ru}_{1-x}\mathrm{Cr}_{x}\mathrm{O}_{2}(001)$ MTJs.
There, the reinforcement of the electron correlation by the Cr substitution supports the altermagnetism of the rutile system.
\par
In this paper, we further explore the TMR effect with Cr-doped $\mathrm{RuO_{2}}$ altermagnetic electrode.
Based on the distinct behaviors between the $\mathrm{RuO_{2}}(001)$ MTJ and the $\mathrm{RuO_{2}}(110)$ MTJ,
we can expect intriguing TMR properties when we use the $\mathrm{Ru}_{1-x}\mathrm{Cr}_{x}\mathrm{O}_{2}(110)$ MTJs.
Hence, we theoretically study the TMR effect in the $\mathrm{Ru}_{1-x}\mathrm{Cr}_{x}\mathrm{O}_{2}(110)/\mathrm{TiO_{2}}(110)/\mathrm{Ru}_{1-x}\mathrm{Cr}_{x}\mathrm{O}_{2}(110)$ tunnel junction.
Using a first-principles approach, we calculate the tunneling transport in the $(110)$-stacked MTJs.
We show that a finite spin-polarization appears in the bulk $\mathrm{Ru}_{1-x}\mathrm{Cr}_{x}\mathrm{O}_{2}(110)$,
which contributes to the emergence of the TMR effect.
The resultant TMR ratio in the $\mathrm{Ru}_{1-x}\mathrm{Cr}_{x}\mathrm{O}_{2}(110)$-based MTJs appear to be larger than the $\mathrm{Ru}_{1-x}\mathrm{Cr}_{x}\mathrm{O}_{2}(001)$-based MTJs depending on the Cr concentration.
As well as the total spin polarization of the electric current flowing through the bulk $\mathrm{Ru}_{1-x}\mathrm{Cr}_{x}\mathrm{O}_{2}(110)$,
we also discuss the role of the momentum-resolved spin splitting for the TMR effect based on the discrepancy of the net spin polarization of electric current and the TMR effect.
\section{System and Method}
\label{sec:method}
\begin{figure}[tbh]
	\centering
	\includegraphics[width=86mm]{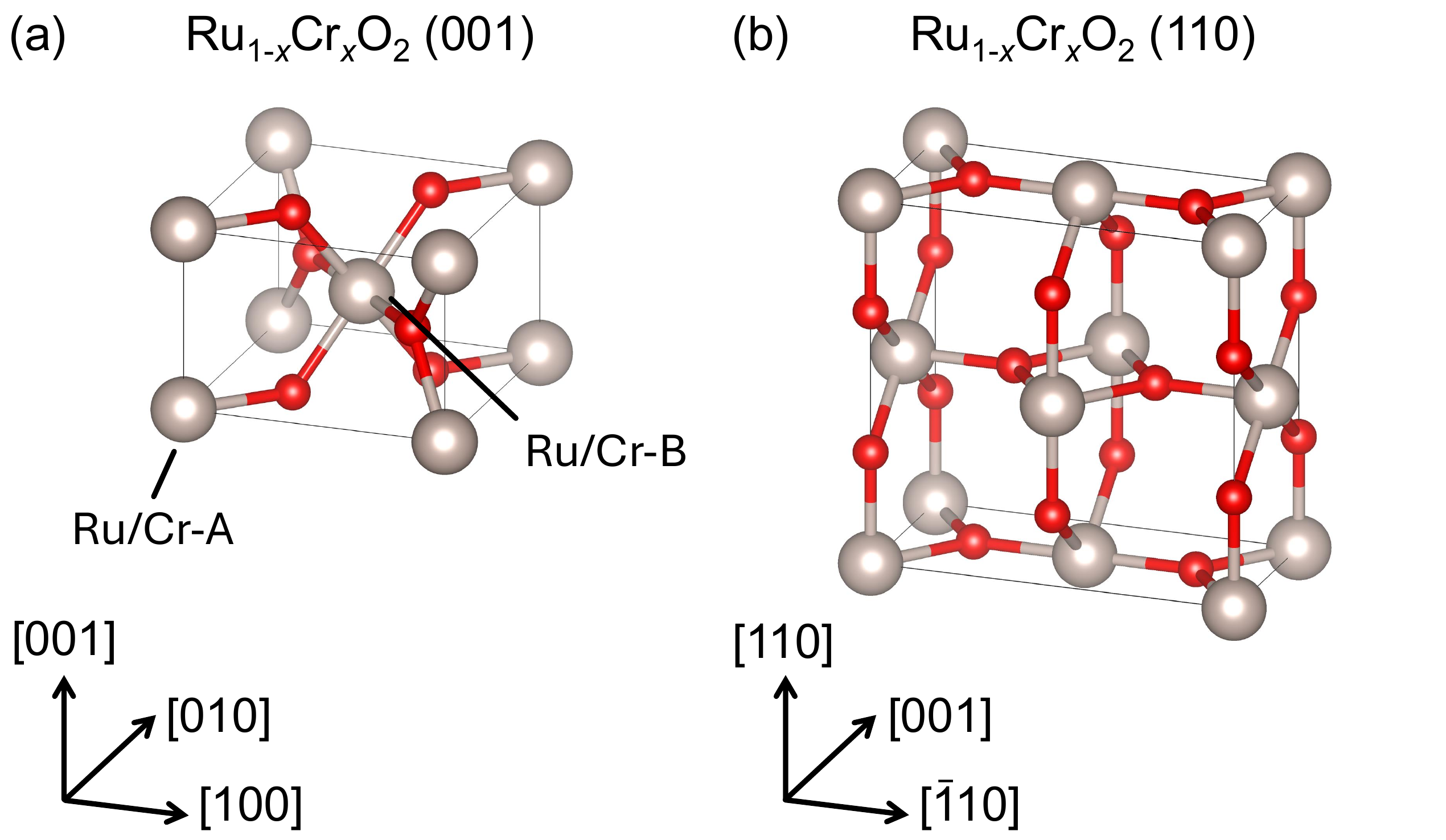}
	\caption{%
		(a) Crystal structures of the rutile $\mathrm{Ru}_{1-x}\mathrm{Cr}_{x}\mathrm{O}_{2}$ primitive cell.
		(b) Crystal structure of the rutile $\mathrm{Ru}_{1-x}\mathrm{Cr}_{x}\mathrm{O}_{2}$ stacked along $[110]$-direction.
	}
	\label{fig:structure}
\end{figure}
As an electrode for the MTJ, we use the rutile $\mathrm{Ru}_{1-x}\mathrm{Cr}_{x}\mathrm{O}_{2}$,
whose primitive cell is shown in Fig.~\ref{fig:structure}(a).
In the altermagnetic state of $\mathrm{Ru}_{1-x}\mathrm{Cr}_{x}\mathrm{O}_{2}$,
the two Ru/Cr-sites, Ru/Cr-A and B, are inequivalent to each other.
The $a$- and $c$-axes lattice constants are respecrively set as $a_{\mathrm{RuO_{2}}} = 4.4919$~\si{\angstrom} and $c_{\mathrm{RuO_{2}}} = 3.1066$~\si{\angstrom} for all compositions.
We stack $\mathrm{Ru}_{1-x}\mathrm{Cr}_{x}\mathrm{O}_{2}$ along the $\left[ 110 \right]$-direction using the supercell shown in Fig.~\ref{fig:structure}(b) to construct the MTJ with rutile $\mathrm{TiO_{2}}(110)$ as the barrier layer.
Namely, we construct the $\mathrm{Ru}_{1-x}\mathrm{Cr}_{x}\mathrm{O}_{2}(110)/\mathrm{TiO_{2}}(110)/\mathrm{Ru}_{1-x}\mathrm{Cr}_{x}\mathrm{O}_{2}(110)$ MTJs.
The in-plane lattice constants of the $\mathrm{TiO_{2}}(110)$ layer are matched to those of $\mathrm{Ru}_{1-x}\mathrm{Cr}_{x}\mathrm{O}_{2}(110)$,
and the out-of-plane lattice constant of $\mathrm{TiO_{2}}$ is $\sqrt{2}a_{\mathrm{TiO_{2}}}$,
where $a_{\mathrm{TiO_{2}}} = 4.5941$~\si{\angstrom} is the $a$-axis lattice constant of the primitive cell of $\mathrm{TiO_{2}}$.
The interfacial distance between $\mathrm{Ru}_{1-x}\mathrm{Cr}_{x}\mathrm{O}_{2}(110)$ and $\mathrm{TiO_{2}}(110)$ is set to be the average of $\sqrt{2}a_{\mathrm{RuO_{2}}}/2$ and $\sqrt{2}a_{\mathrm{TiO_{2}}}/2$.
\par
We calculate first-principles electronic structures based on the density functional theory (DFT)~\cite{Hohenberg1964_PhysRev_136_B864,Kohn1965_PhysRev_140_A1133} with the \textsc{Quantum ESPRESSO} (QE) package~\cite{Giannozzi2009_JPhysCondensMatter_21_395502,Giannozzi2017_JPhysCondensMatter_29_465901}.
The exchange correlation is incorporated by the generalized gradient approximation with the Perdew--Berke--Ernzerhof form~\cite{Perdew1996_PhysRevLett_77_3865}.
The norm-conserved pseudopotential taken from \text{PseudoDojo}~\cite{DalCorso2014_ComptMaterSci_95_337} is used.
The energy cutoffs are respectively set as 110~Ry and 440~Ry for the wave-functions and charge density.
We employ the virtual crystal approximation to substitute Ru with Cr,
where we mix the potential of Ru and Cr and generate the potential of $\mathrm{Ru}_{1-x}\mathrm{Cr}_{x}$~\cite{Nordheim1931_AnnPhys_401_607,Bellaiche2000_PhysRevB_61_7877}.
\par
We calculate the TMR effect employing the scattering theory method~\cite{Choi1999_PhysRevB_59_2267,Smogunov2004_PhysRevB_70_045417} combined with the Landauer--B\"{u}ttiker formula~\cite{Landauer1957_IBMJResDev_1_3,Landauer1970_PhilMag_21_863,Buttiker1986_PhysRevLett_57_1761,Buttiker1988_IBMJResDevelop_32_317}.
We perform the self-consistent field (scf) calculations for each of the lead and the scattering region, 
and attach them to construct the MTJ and calculate the transmission.
The lead part consists of the (110)-faced $\mathrm{Ru}_{1-x}\mathrm{Cr}_{x}\mathrm{O}_{2}$ shown in Fig.~\ref{fig:structure}(b),
and the scattering region consists of the $\mathrm{TiO_{2}}(110)$ barrier layers and a few layers of $\mathrm{Ru}_{1-x}\mathrm{Cr}_{x}\mathrm{O}_{2}(110)$ attached to both sides of $\mathrm{TiO_{2}}(110)$.
The $\boldsymbol{k}$-mesh of the scf calculation is $10 \times 20 \times 10$ for the lead and $10 \times 20 \times 1$ for the scattering region.
To smoothly connect the scattering region and the lead,
we double the supercell of the scattering region in the scf calculation for the antiparallel configuration,
which is cut into half when we calculate the transmission.
Here, the parallel and antiparallel configurations of the MTJs are defined by the relative direction of the magnetic moments in the same Ru/Cr sublattices in the two distinct electrodes;
the MTJ has the parallel (antiparallel) configuration when the magnetic moments in the same Ru/Cr sublattice in the two electrodes align parallelly (antiparallelly).
\par
The total transmission, $T_{\text{tot}}(E)$, at the energy, $E$, is calculated as
\begin{align}
	T_{\text{tot}}(E) = \dfrac{1}{N_{\boldsymbol{k}_{\parallel}}} \sum_{\boldsymbol{k}_{\parallel}, \sigma = \uparrow, \downarrow} T_{\sigma}(\boldsymbol{k}_{\parallel}, E).
	\label{eq:transmission}
\end{align}
Here, $N_{\boldsymbol{k}_{\parallel}}$ is the number of total $\boldsymbol{k}_{\parallel} = (k_{x}, k_{y})$ points, 
and $T_{\sigma}(\boldsymbol{k}_{\parallel}, E)$ is the partial transmission of the spin-$\sigma$ channel at the $\boldsymbol{k}_{\parallel}$-point with the energy $E$.
We take $N_{\boldsymbol{k}_{\parallel}} = 151 \times 151$ at $E = E_{\mathrm{F}}$ at each concentration for $N_{\mathrm{TiO_{2}}(110)} = 4$,
where $N_{\mathrm{TiO_{2}}(110)}$ is the number of $\mathrm{TiO_{2}(110)}$ barrier layers in the MTJ.
For the other cases, we take $N_{\boldsymbol{k}_{\parallel}} = 101 \times 101$.
\par
Additionally, we calculate the number of conduction channels at each $\boldsymbol{k}_{\parallel}$-point and the effective polarization of the bulk $\mathrm{Ru}_{1-x}\mathrm{Cr}_{x}\mathrm{O}_{2}(110)$.
The momentum dependence of the transport spin polarization is given as~\cite{Shao2021_NatCommun_12_7061}
\begin{align}
	p(\boldsymbol{k}_{\parallel})
= \dfrac{N_{\uparrow}(\boldsymbol{k}_{\parallel}) - N_{\downarrow}(\boldsymbol{k}_{\parallel})}
				{N_{\uparrow}(\boldsymbol{k}_{\parallel}) + N_{\downarrow}(\boldsymbol{k}_{\parallel})}.
	\label{eq:polarization_kdep}
\end{align}
Here, $N_{\uparrow/\downarrow}(\boldsymbol{k}_{\parallel})$ is the number of conduction channels of the up/down spins at $\boldsymbol{k}_{\parallel}$.
The total spin polarization of the current, $p_{\text{tot}}$, is calculated as
\begin{align}
	p_{\text{tot}}
= \dfrac{\displaystyle\sum_{\boldsymbol{k}_{\parallel}}\left( N_{\uparrow}(\boldsymbol{k}_{\parallel}) - N_{\downarrow}(\boldsymbol{k}_{\parallel}) \right)}
				{\displaystyle\sum_{\boldsymbol{k}_{\parallel}}\left( N_{\uparrow}(\boldsymbol{k}_{\parallel}) + N_{\downarrow}(\boldsymbol{k}_{\parallel}) \right)}.
	\label{eq:polarization_total}
\end{align}
The $\boldsymbol{k}_{\parallel}$-mesh is $151\times 151$ at $E = E_{\mathrm{F}}$, and $101 \times 101$ for the other cases.
\par
To examine the electronic states of the barrier region, we also calculate the projected density of states (PDOS) onto each atomic orbital for the scattering region in the parallel configuration.
Before the calculation of the PDOS, we perform the non-scf calculation with $16\times 32 \times 1$ $\boldsymbol{k}$-points after the scf calculation described above.
\section{Results and discussions}
\label{sec:results}
\subsection{Spin polarization of bulk $\mathrm{Ru}_{1-x}\mathrm{Cr}_{x}\mathrm{O}_{2}$}
\begin{figure}[tbh]
	\centering
	\includegraphics[width=86mm]{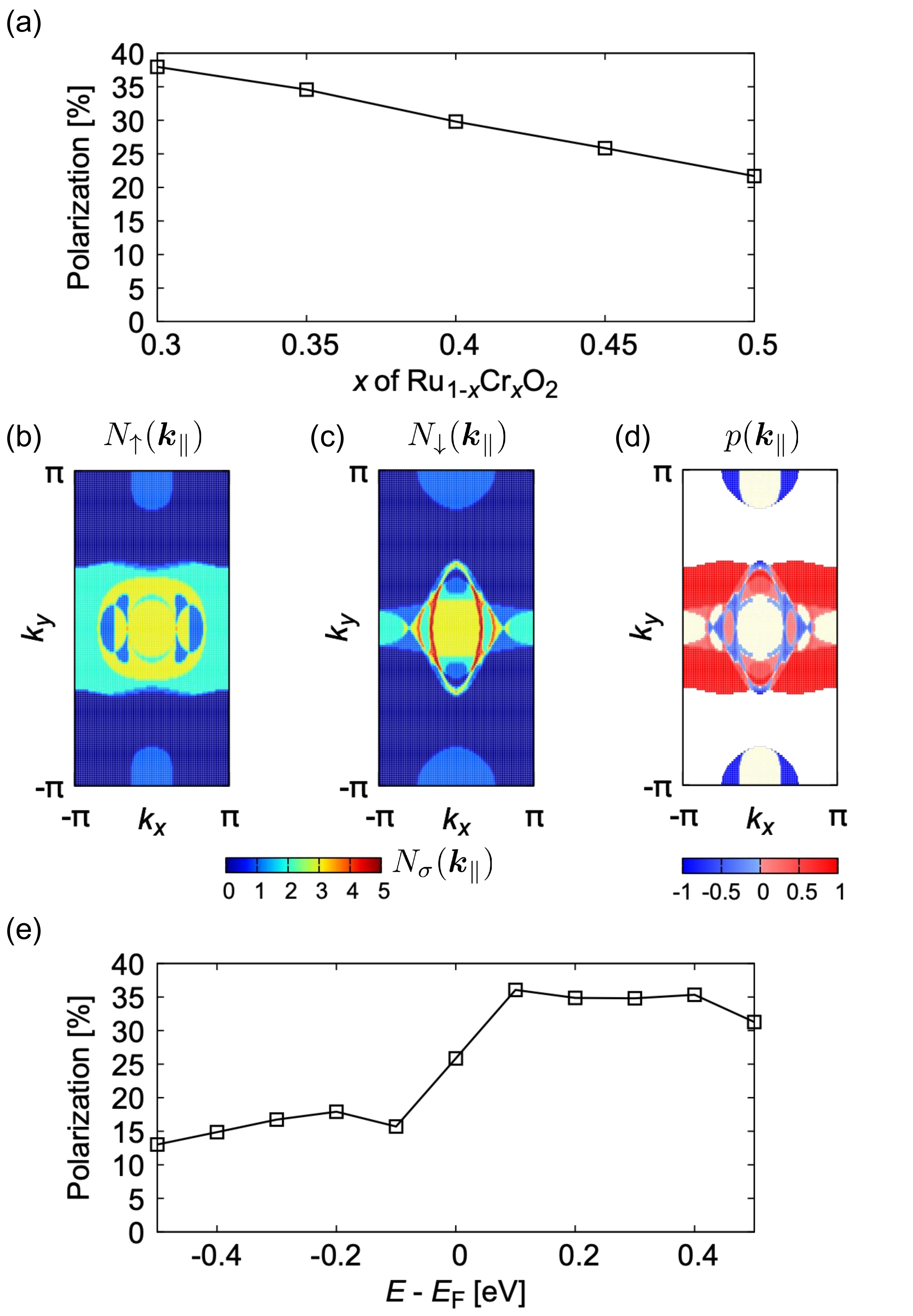}
	\caption{%
		Calculation results of the spin polarization in the bulk $\mathrm{Ru}_{1-x}\mathrm{Cr}_{x}\mathrm{O}_{2}(110)$ systems.
		(a) Cr-concentration dependence of the polarization $p_{\mathrm{tot}}$ (Eq.~(\ref{eq:polarization_total})) at $E = E_{\mathrm{F}}$.
		(b), (c) In-plane momentum dependence of the number of conduction channels for (b) up and (c) down spins at $E = E_{\mathrm{F}}$ for the $x = 0.45$ system.
		(d) In-plane momentum dependence of the polarization, $p(\boldsymbol{k}_{\parallel})$ (Eq.~(\ref{eq:polarization_kdep})), corresponding to (b) and (c).
		(e) Energy dependence of $p_{\mathrm{tot}}$ for the $x = 0.45$ system.
				Note that the $\boldsymbol{k}$-mesh is $151 \times 151$ at $E = E_{\mathrm{F}}$ and $101 \times 101$ for others.
	}
	\label{fig:channel_polarization}
\end{figure}
We discuss the transport spin polarization in the bulk $\mathrm{Ru}_{1-x}\mathrm{Cr}_{x}\mathrm{O}_{2}(110)$.
In Fig.~\ref{fig:channel_polarization}(a), we show the Cr concentration dependence of the total spin polarization calculated by Eq.~(\ref{eq:polarization_total}).
We find that the total spin polarization is finite in all cases,
and the size of the polarization decreases as $x$ increases.
The distributions of the numbers of the conduction channels of up- and down-spins in the momentum space at $E = E_{\mathrm{F}}$ for $x = 0.45$ are shown in Figs.~\ref{fig:channel_polarization}(b) and \ref{fig:channel_polarization}(c), respectively.
The numbers of conduction channels are large around the $\Gamma$-point for both of up and down spins.
Based on these distributions, 
we calculate the momentum dependence of the polarization $p(\boldsymbol{k}_{\parallel})$ (Eq.~(\ref{eq:polarization_kdep})),
which is shown in Fig.~\ref{fig:channel_polarization}(d).
We see that the in-plane spin polarization appears not at the $\Gamma$-point but the area around $\boldsymbol{k}_{\parallel} = 0$.
\par
The energy dependence of the total polarization $p_{\mathrm{tot}}$ for the $x = 0.45$ system is shown in Fig.~\ref{fig:channel_polarization}(e).
The sign of total polarizaton is always positive from $E - E_{\mathrm{F}} = -0.5$~eV to $0.5$~eV,
and its magnitude shows a tendency to increase as the chemical potential increases, reaching around 35\% at the largest.
\subsection{Tunnel magnetoresistance effect}
\begin{figure*}[tbh]
	\centering
	\includegraphics[width=172mm]{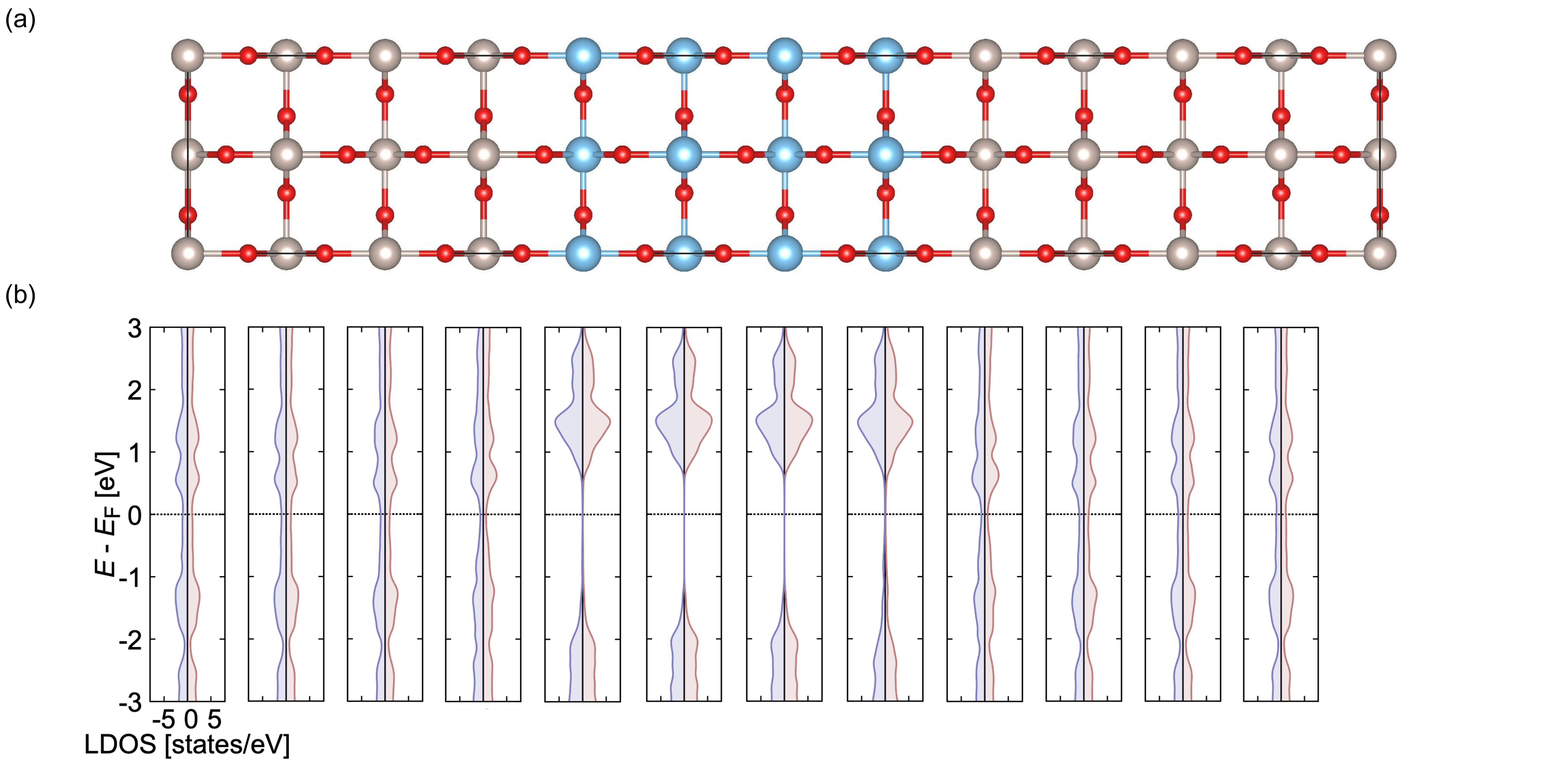}
	\caption{%
		(a) Crystal structure of the $\mathrm{Ru}_{1-x}\mathrm{Cr}_{x}\mathrm{O}_{2}(110)/\mathrm{TiO_{2}}(110)/\mathrm{Ru}_{1-x}\mathrm{Cr}_{x}\mathrm{O}_{2}(110)$ tunnel junction with four monolayers of $\mathrm{TiO_{2}}(110)$.
		Specifically, the scattering region used in the calculations is shown.
		(b) Density of states (DOS) of the scattering region with the parallel configuration for the $x = 0.45$ system.
		Each DOS contains the projected DOS of the layers with Ru/Ti atoms and the oxygen atoms nearby.
	}
	\label{fig:layerdos}
\end{figure*}
Next, we discuss tunneling transport and the TMR effect in the $\mathrm{Ru}_{1-x}\mathrm{Cr}_{x}\mathrm{O}_{2}(110)/\mathrm{TiO_{2}}(110)/\mathrm{Ru}_{1-x}\mathrm{Cr}_{x}\mathrm{O}_{2}(110)$ MTJs.
Figure~\ref{fig:layerdos}(a) shows the crystal structure of the scattering region used in the TMR calculations with $N_{\mathrm{TiO_{2}}} = 4$.
The energy dependence of the density of states (DOS) of each layer of the scattering region is shown in Fig.~\ref{fig:layerdos}(b) for the parallel configuration of the $x = 0.45$ system.
We see that the DOS is small enough around $E = E_{\mathrm{F}}$ at the $\mathrm{TiO_{2}}(110)$ barrier layers.
This ensures that we can indeed discuss the tunneling transport in this system.
\begin{figure}[tbh]
	\centering
	\includegraphics[width=86mm]{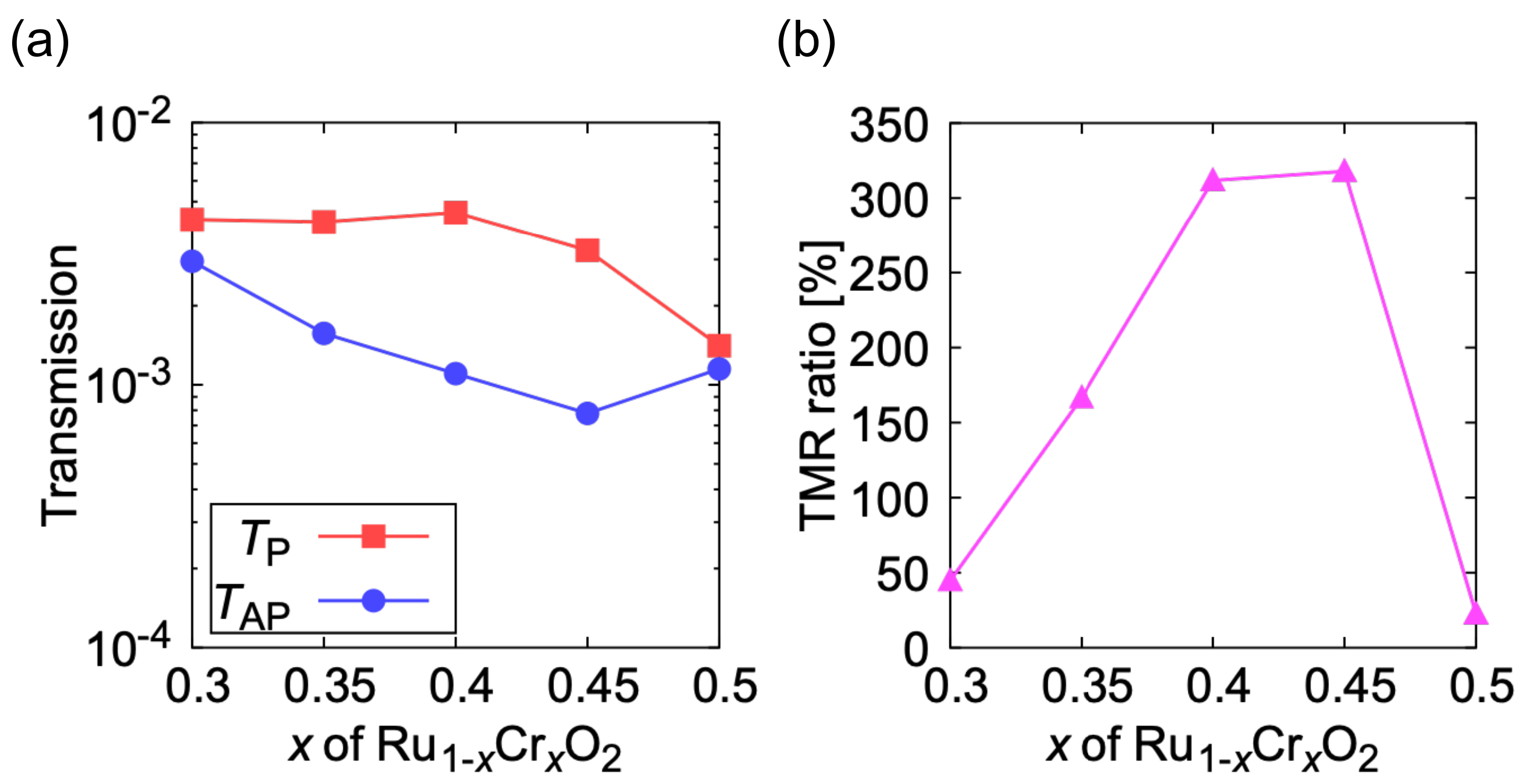}
	\caption{%
		(a) Cr-doping amount dependence of the total transmission for the parallel and antiparallel configurations, $T_{\mathrm{P}}$ and $T_{\mathrm{AP}}$, respectively, at $E = E_{\text{F}}$ with four monolayers of $\mathrm{TiO_{2}}(110)$.
		(b) Cr-doping amount dependence of the TMR ratio calculated by $\left( T_{\mathrm{P}} - T_{\mathrm{AP}} \right) / T_{\mathrm{AP}} \times 100\%$ corresponding to (a).
	}
	\label{fig:TMR_Crdep}
\end{figure}
\par
In Fig.~\ref{fig:TMR_Crdep}(a), we show the Cr-concentration dependence of total transmission at $E = E_{\mathrm{F}}$, $T_{\text{tot}}(E_{\mathrm{F}})$, for the parallel and antiparallel configurations, $T_{\mathrm{P}}$ and $T_{\mathrm{AP}}$, respectively.
Overall, $T_{\mathrm{P}}$ shows a downward trend as the Cr-concentration increases to $x = 0.50$.
By contrast, $T_{\mathrm{AP}}$ decreases with increasing Cr-concentration and takes a minimum at $x = 0.45$.
Then, the TMR ratio, calculated by $[\text{TMR ratio}~[\%]] = (T_{\text{P}}-T_{\text{AP}})/T_{\text{AP}}\times 100$, 
increases as $x$ increases and peaks at $x = 0.45$, reaching around 300\%.
In the $\mathrm{Ru}_{1-x}\mathrm{Cr}_{x}\mathrm{O}_{2}(001)/\mathrm{TiO_{2}}(001)/\mathrm{Ru}_{1-x}\mathrm{Cr}_{x}\mathrm{O}_{2}(001)$ MTJs,
the TMR ratio is around 100\%--200\%~\cite{Tanaka2024_PhysRevB_110_064433},
so these results suggest that we can enhance the TMR effect with the $(110)$-stacking when we properly set the Cr-concentration.
\begin{figure}[tbh]
	\centering
	\includegraphics[width=86mm]{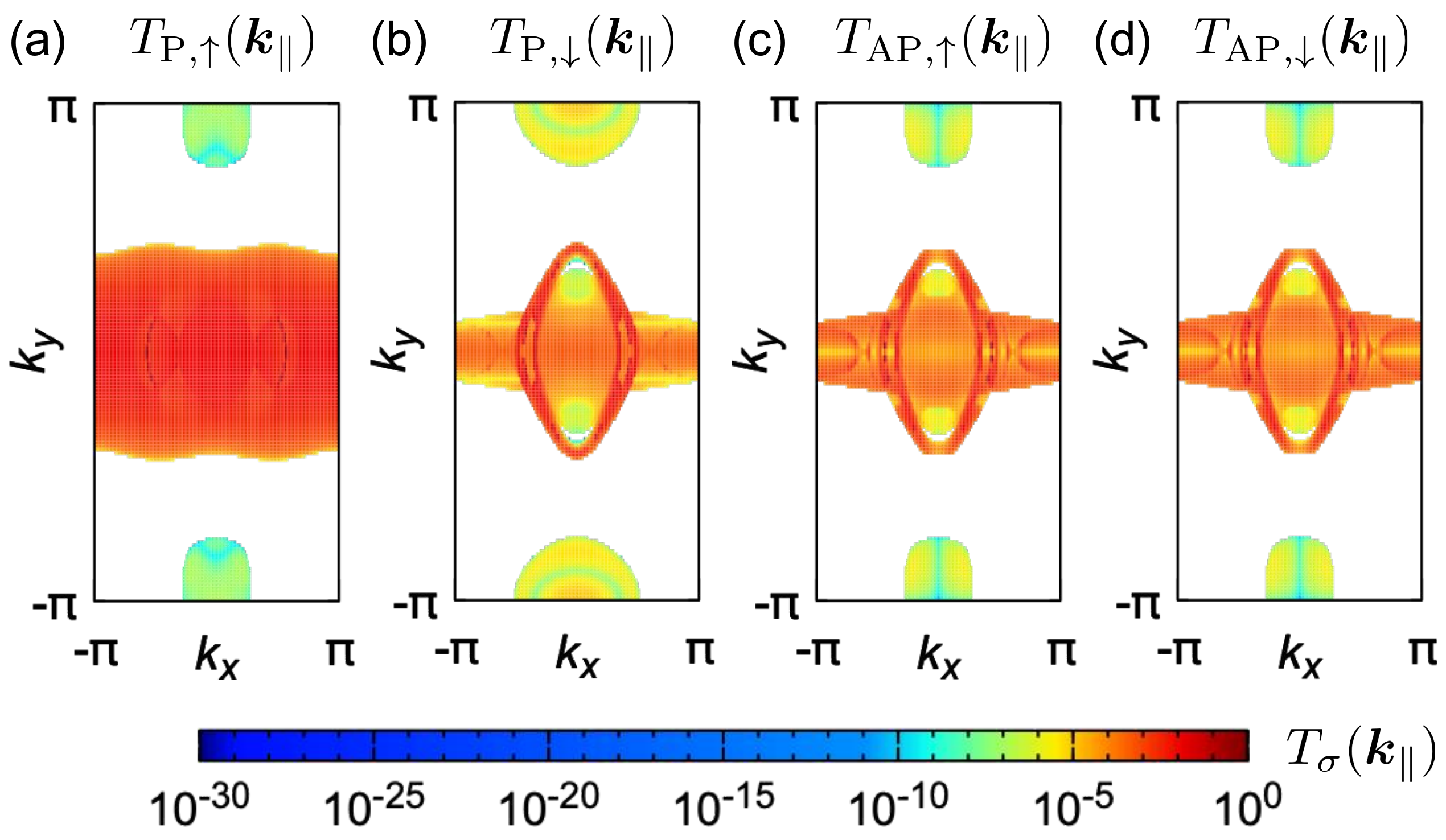}
	\caption{%
		Momentum resolved transmission at $E = E_{\text{F}}$ for the $x = 0.45$ system with four monolayers of $\mathrm{TiO_{2}}(110)$.
		(a), (b) Transmission for the parallel configuration with (a) up and (b) down spins.
		(c), (d) Transmission for the antiparallel configuration with (c) up and (d) down spins.
	}
	\label{fig:TMR_kdep}
\end{figure}
\par
The momentum dependence of the partial transmission at $E = E_{\mathrm{F}}$ for the $x = 0.45$ system is shown in Figs.~\ref{fig:TMR_kdep}(a)--(d).
For the parallel configuration, the transmission distributions directly reflect those of the conduction channels shown in Figs.~\ref{fig:channel_polarization}(b) and \ref{fig:channel_polarization}(c).
Namely, the transmission can be nonzero in the regions where the number of conduction channels is finite, for each of the up- and down-spin channels.
By contrast, for the antiparallel configuration, the transmission distribution is determined by the combination of the up- and down-spin channels,
due to the reversal of the magnetic moments in one-side of the $\mathrm{Ru}_{1-x}\mathrm{Cr}_{x}\mathrm{O}_{2}$ electrode.
Thus, the partial transmission takes a finite value at the $\boldsymbol{k}_{\parallel}$ point where both $N_{\uparrow}(\boldsymbol{k}_{\parallel})$ and $N_{\downarrow}(\boldsymbol{k}_{\parallel})$ are nonzero, 
as shown in Figs.~\ref{fig:TMR_kdep}(c) and \ref{fig:TMR_kdep}(d).
\par
We note that the Cr amount dependence of the TMR ratio has the peak structure in the range between $0.3 \leq x \leq 0.5$,
while the total spin polarization $p_{\text{tot}}$ (Eq.~(\ref{eq:polarization_total})) shows the monotonic decrease as $x$ increases.
This discrepancy indicates that the TMR effect cannot be understood only by the total spin polarization of electric current through magnetic electrodes as a bulk material, 
but by other essential components like the momentum dependent spin-splitting or the interfacial spin polarization between the magnetic electrodes and the barrier.
\begin{figure}[tbh]
	\centering
	\includegraphics[width=86mm]{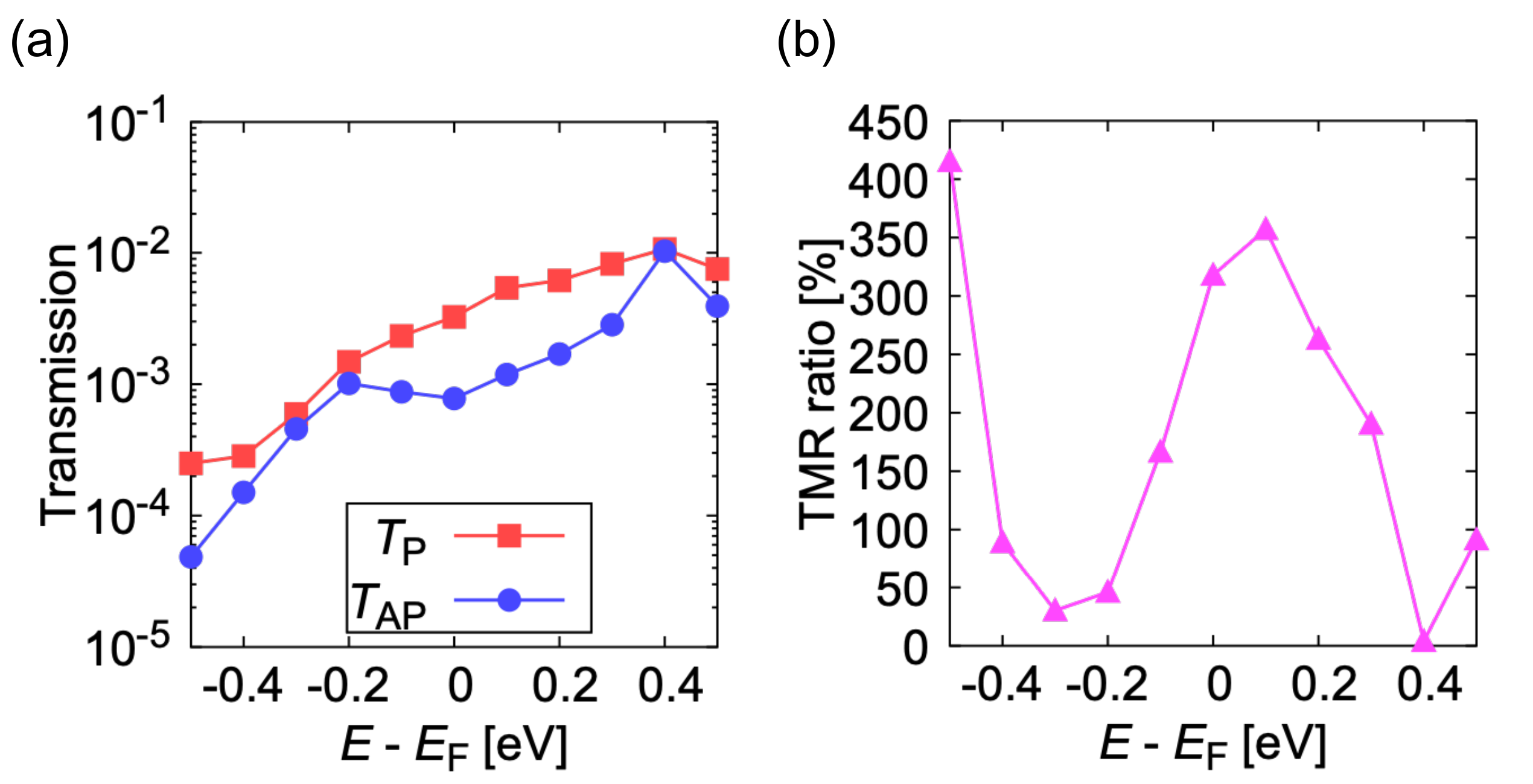}
	\caption{%
		(a) Energy dependence of the total transmission for the parallel and antiparallel configurations, $T_{\text{P}}$ and $T_{\text{AP}}$, respectively, for the $x = 0.45$ system with four monolayers of $\mathrm{TiO_{2}}$.
		(b) Energy dependence of the TMR ratio corresponding to (a).
		Note that the $\boldsymbol{k}$-mesh is $151 \times 151$ at $E = E_{\mathrm{F}}$ and $101 \times 101$ for others.
	}
	\label{fig:TMR_Edep}
\end{figure}
\par
We also calculate the energy dependence of the total transmission and the TMR ratio for the $x = 0.45$ system.
In Fig.~\ref{fig:TMR_Edep}(a), we show the energy dependence of the total transmission.
The total transmission tends to increase as $E - E_{\mathrm{F}}$ increases,
which is because we go away from the center of the band gap and approach its edge by increasing the energy, as indicated in Fig.~\ref{fig:layerdos}(b).
The corresponding TMR ratio is shown in Fig~\ref{fig:TMR_Edep}(b).
The TMR ratio highly depends on the energy;
it increases to $\sim 400\%$ at $E = E_{\mathrm{F}} - 0.5$~eV and to $\sim 350\%$ at $E = E_{\mathrm{F}} + 0.1$~eV,
while it takes almost zero at $E = E_{\mathrm{F}} + 0.4$~eV.
Similarly to the discussion of the dependence of the Cr-doping amount, 
the energy dependence of the TMR ratio does not correspond to that of the total spin polarization shown in Fig.~\ref{fig:channel_polarization}(e),
where $p_{\mathrm{tot}}$ shows a monotonic decrease when the chemical potential increases.
This again indicates that the TMR effect is determined not only by the total spin polarization of the bulk magnetic properties of the electrodes, 
but also other effects such as the momentum dependent spin polarization of the magnetic electrodes or the spin polarization at the interface,
as is also indicated in Ref.~\cite{Jiang2023_PhysRevB_108_174439} for the MTJ with pristine $\mathrm{RuO_{2}}(110)$.
\begin{figure}[tbh]
	\centering
	\includegraphics[width=86mm]{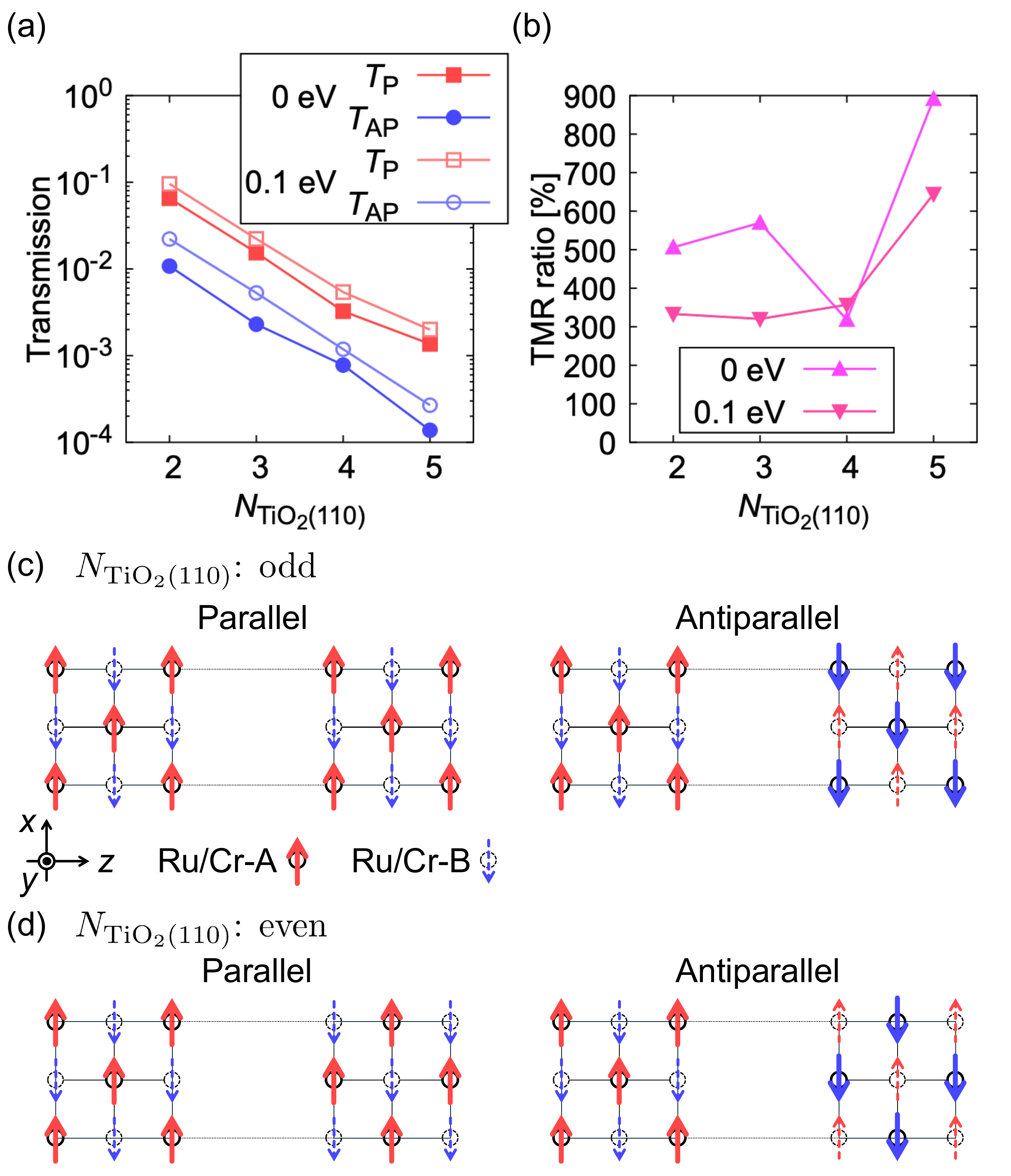}
	\caption{%
		(a) $\mathrm{TiO_{2}}(110)$ barrier thickness dependence of the total transmission for the parallel and antiparallel configurations, $T_{\text{P}}$ and $T_{\text{AP}}$, respectively, at $E - E_{\mathrm{F}} = 0$~eV and $0.1$~eV for the $x = 0.45$ system.
		(b) TMR ratio corresponding to (a).
		Note that the $\boldsymbol{k}$-mesh is $151 \times 151$ at $E = E_{\mathrm{F}}$ for $N_{\mathrm{TiO_{2}}(110)} = 4$ and $101 \times 101$ for all other cases.
		(c), (d) Schematic illustrations of the parallel and antiparallel configurations of the $\mathrm{Ru}_{1-x}\mathrm{Cr}_{x}\mathrm{O}_{2}(110)$-based magnetic tunnel junctions with the number of layers (c) odd and (d) even.
		Arrows represent the magnetic moments carried by the Ru/Cr ions.
		Circles and arrows with solid lines correspond to the Ru/Cr-A site and its magentic moment respectively, 
		and those with broken lines correspond to the Ru/Cr-B site and its magentic moment.
		Namely, the objects with solid lines have the same $y$-coordinates,
		and those with broken lines also have the same $y$-coordinates, 
		while the objects with the solid and broken lines have different $y$-coordinates by $a_{\mathrm{RuO_{2}}}/2$.
	}
	\label{fig:TMR_layerdep}
\end{figure}
\par
Additionally, we calculate the TMR effect in MTJs with $N_{\mathrm{TiO_{2}}(110)} \neq 4$.
Figure~\ref{fig:TMR_layerdep}(a) shows the barrier thickness dependence of $T_{\mathrm{P}}$ and $T_{\mathrm{AP}}$ at $E - E_{\mathrm{F}} = 0$~eV and $0.1$~eV for the case of $x = 0.45$.
Both $T_{\mathrm{P}}$ and $T_{\mathrm{AP}}$ exponentially decrease as the thickness of the $\mathrm{TiO_{2}}(110)$ barrier increases,
indicating the tunneling transport, in addition to the DOS of the scattering region shown in Fig.~\ref{fig:layerdos}(b).
The barrier thickness dependence of the TMR ratio corresponding to Fig.~\ref{fig:TMR_layerdep}(a) is shown in Fig.~\ref{fig:TMR_layerdep}(b).
At $E = E_{\mathrm{F}}$, the TMR ratio shows an oscillatory behavior with respect to $N_{\mathrm{TiO_{2}}(110)}$,
while such an oscillation is not observed at $E = E_{\mathrm{F}}+0.1$~eV.
\par
This behavior can be understood by taking account of the effect of the interfacial magnetic structures of the MTJ.
We schematically show the sets of the parallel and antiparallel configurations of the MTJ with $N_{\mathrm{TiO_{2}}(110)}$ odd and even in Figs.~\ref{fig:TMR_layerdep}(c) and \ref{fig:TMR_layerdep}(d), respectively.
When $N_{\mathrm{TiO_{2}}(110)}$ is odd, the magnetic moments in the same sublattices face each other across the $\mathrm{TiO_{2}}(110)$ barrier with the same in-plane ($xy$) coordinates.
These magnetic moments have the parallel (antiparallel) directions for the parallel (antiparallel) configuration of the MTJ as shown in Fig.~\ref{fig:TMR_layerdep}(c).
On the other hand, when $N_{\mathrm{TiO_{2}}(110)}$ is even,
the magnetic moments in the same sublattices face each other with different in-plane coordinates;
the interfacial magnetic moments in the same sublattices misalign as described in Fig.~\ref{fig:TMR_layerdep}(d).
Hence, when $N_{\mathrm{TiO_{2}}(110)}$ is odd, the TMR effect can be pronounced due to the matching of the magnetic configurations, which may enhance the TMR ratio for the odd-$N_{\mathrm{TiO_{2}}(110)}$ systems compared with the even-$N_{\mathrm{TiO_{2}}(110)}$ systems,
thereby exhibiting an oscillatory behavior of the TMR ratio.
However, as shown for the $E = E_{\mathrm{F}}+0.1$~eV case, such an enhancement of the TMR ratio does not always occur for the MTJs with odd $N_{\mathrm{TiO_{2}}(110)}$;
the TMR ratio is determined by the balance between the interfacial magnetic structures and other effects such as the total spin-polarization of the magnetic electrode as the bulk or the momentum-dependent spin-polarization.
We note that similar discussions on the relation between the TMR ratio and the interfacial magnetic configurations can be found in Ref.~\cite{Tanaka2023_PhysRevB_107_214442} with the lattice model calculation and Ref.~\cite{Jiang2023_PhysRevB_108_174439} with the DFT calculation in the $\mathrm{RuO_{2}}(110)/\mathrm{TiO_{2}}(110)/\mathrm{RuO_{2}}(110)$ MTJ.
Particularly, in Ref.~\cite{Tanaka2023_PhysRevB_107_214442},
it has been found that even the sign reversal of the TMR ratio can occur originating from the interfacial structures in the two-sublattice ferrimagnetic MTJs,
where there are two types of interfacial configurations reflecting the two-sublattice magnetic structure.
In one configuration, the TMR ratio is positive in principle. 
In the other configuration with common interaction parameters, by contrast, the TMR ratio can be negative, depending on the energy.
Given that each of these two interfacial magnetic configurations correspond to one of the present $\mathrm{Ru}_{1-x}\mathrm{Cr}_{x}\mathrm{O}_{2}(110)$-MTJs with even or odd $N_{\mathrm{TiO_{2}}(110)}$,
the oscillatory behavior shown in Fig.~\ref{fig:TMR_layerdep}(b) can be regarded as the sign change of the TMR ratio in the lattice model between two different configurations.
Namely, as the TMR ratio does not always change its sign in the lattice model calculations,
it should be reasonable that the presence or absence of the oscillation of the TMR ratio depends on the energy in the $\mathrm{Ru}_{1-x}\mathrm{Cr}_{x}\mathrm{O}_{2}(110)$-based MTJs.
\section{Summary}
\label{sec:summary}
In summary, we have studied the tunnel magnetoresistance (TMR) effect with a doped altermagnet $\mathrm{Ru}_{1-x}\mathrm{Cr}_{x}\mathrm{O}_{2}(110)$ using first-principles calculations.
We have shown that a finite spin-polarization appears in the $\mathrm{Ru}_{1-x}\mathrm{Cr}_{x}\mathrm{O}_{2}(110)$ structure,
which can support a finite TMR effect in the $\mathrm{Ru}_{1-x}\mathrm{Cr}_{x}\mathrm{O}_{2}(110)$-based magnetic tunnel junctions (MTJs).
We have constructed the $\mathrm{Ru}_{1-x}\mathrm{Cr}_{x}\mathrm{O}_{2}(110)/\mathrm{TiO_{2}}(110)/\mathrm{Ru}_{1-x}\mathrm{Cr}_{x}\mathrm{O}_{2}(110)$ MTJ, calculated the transmission, and shown that a finite TMR effect actually appears.
The results of the calculations have suggested that the TMR effect is generated by the momentum-dependent spin splitting and the interfacial magnetic structure, along with the spin-polarization possessed by the bulk $\mathrm{Ru}_{1-x}\mathrm{Cr}_{x}\mathrm{O}_{2}(110)$.
\begin{acknowledgments}
This work was supported by JST-Mirai Program (Grant No.~JPMJMI20A1), JST-ASPIRE (Grant No. JPMJAP2317), JST-CREST (Grant No. JPMJCR23O4), JSPS-KAKENHI (No.~24K00581, No.~JP25K17935, No.~JP25K24713, No.~JP25K21684, No.~JP25H01252, No.~JP26H02010), and the RIKEN TRIP initiative (RIKEN Quantum, AGIS, Many-body Electron Systems).
Part of the calculations in this study has been done using the ISSP Supercomputer.
We use the \textsc{VESTA}~\cite{Momma2011_JApplCryst_44_1272} and \textsc{Atomsk}~\cite{Hirel2015ComputPhysCommun197_212} softwares for the visualization of the crystal structures.
\end{acknowledgments}
\input{arXiv.bbl}
\end{document}

%% file: arXiv.bbl
%

%% file: arXiv.bbl
\begin{thebibliography}{77}%
\makeatletter
\providecommand \@ifxundefined [1]{%
 \@ifx{#1\undefined}
}%
\providecommand \@ifnum [1]{%
 \ifnum #1\expandafter \@firstoftwo
 \else \expandafter \@secondoftwo
 \fi
}%
\providecommand \@ifx [1]{%
 \ifx #1\expandafter \@firstoftwo
 \else \expandafter \@secondoftwo
 \fi
}%
\providecommand \natexlab [1]{#1}%
\providecommand \enquote  [1]{``#1''}%
\providecommand \bibnamefont  [1]{#1}%
\providecommand \bibfnamefont [1]{#1}%
\providecommand \citenamefont [1]{#1}%
\providecommand \href@noop [0]{\@secondoftwo}%
\providecommand \href [0]{\begingroup \@sanitize@url \@href}%
\providecommand \@href[1]{\@@startlink{#1}\@@href}%
\providecommand \@@href[1]{\endgroup#1\@@endlink}%
\providecommand \@sanitize@url [0]{\catcode `\\12\catcode `\$12\catcode
  `\&12\catcode `\#12\catcode `\^12\catcode `\_12\catcode `\%12\relax}%
\providecommand \@@startlink[1]{}%
\providecommand \@@endlink[0]{}%
\providecommand \url  [0]{\begingroup\@sanitize@url \@url }%
\providecommand \@url [1]{\endgroup\@href {#1}{\urlprefix }}%
\providecommand \urlprefix  [0]{URL }%
\providecommand \Eprint [0]{\href }%
\providecommand \doibase [0]{https://doi.org/}%
\providecommand \selectlanguage [0]{\@gobble}%
\providecommand \bibinfo  [0]{\@secondoftwo}%
\providecommand \bibfield  [0]{\@secondoftwo}%
\providecommand \translation [1]{[#1]}%
\providecommand \BibitemOpen [0]{}%
\providecommand \bibitemStop [0]{}%
\providecommand \bibitemNoStop [0]{.\EOS\space}%
\providecommand \EOS [0]{\spacefactor3000\relax}%
\providecommand \BibitemShut  [1]{\csname bibitem#1\endcsname}%
\let\auto@bib@innerbib\@empty
\bibitem [{\citenamefont {Julliere}(1975)}]{Julliere1975_PhysLettA_54A_225}%
  \BibitemOpen
  \bibfield  {author} {\bibinfo {author} {\bibfnamefont {M.}~\bibnamefont
  {Julliere}},\ }\bibfield  {title} {\bibinfo {title} {{Tunneling between
  ferromagnetic films}},\ }\href
  {https://doi.org/https://doi.org/10.1016/0375-9601(75)90174-7} {\bibfield
  {journal} {\bibinfo  {journal} {Phys. Lett. A}\ }\textbf {\bibinfo {volume}
  {54}},\ \bibinfo {pages} {225--226} (\bibinfo {year} {1975})}\BibitemShut
  {NoStop}%
\bibitem [{\citenamefont {Miyazaki}\ and\ \citenamefont
  {Tezuka}(1995)}]{Miyazaki1995_JMagnMagnMater_139_L231}%
  \BibitemOpen
  \bibfield  {author} {\bibinfo {author} {\bibfnamefont {T.}~\bibnamefont
  {Miyazaki}}\ and\ \bibinfo {author} {\bibfnamefont {N.}~\bibnamefont
  {Tezuka}},\ }\bibfield  {title} {\bibinfo {title} {{Giant magnetic tunneling
  effect in $\mathrm{Fe/Al_{2}O_{3}/Fe}$ junction}},\ }\href
  {https://doi.org/https://doi.org/10.1016/0304-8853(95)90001-2} {\bibfield
  {journal} {\bibinfo  {journal} {J. Magn. Magn. Mater.}\ }\textbf {\bibinfo
  {volume} {139}},\ \bibinfo {pages} {L231--L234} (\bibinfo {year}
  {1995})}\BibitemShut {NoStop}%
\bibitem [{\citenamefont {Moodera}\ \emph {et~al.}(1995)\citenamefont
  {Moodera}, \citenamefont {Kinder}, \citenamefont {Wong},\ and\ \citenamefont
  {Meservey}}]{Moodera1995_PhysRevLett_74_3273}%
  \BibitemOpen
  \bibfield  {author} {\bibinfo {author} {\bibfnamefont {J.~S.}\ \bibnamefont
  {Moodera}}, \bibinfo {author} {\bibfnamefont {L.~R.}\ \bibnamefont {Kinder}},
  \bibinfo {author} {\bibfnamefont {T.~M.}\ \bibnamefont {Wong}},\ and\
  \bibinfo {author} {\bibfnamefont {R.}~\bibnamefont {Meservey}},\ }\bibfield
  {title} {\bibinfo {title} {{Large Magnetoresistance at Room Temperature in
  Ferromagnetic Thin Film Tunnel Junctions}},\ }\href
  {https://doi.org/10.1103/PhysRevLett.74.3273} {\bibfield  {journal} {\bibinfo
   {journal} {Phys. Rev. Lett.}\ }\textbf {\bibinfo {volume} {74}},\ \bibinfo
  {pages} {3273--3276} (\bibinfo {year} {1995})}\BibitemShut {NoStop}%
\bibitem [{\citenamefont {Butler}\ \emph {et~al.}(2001)\citenamefont {Butler},
  \citenamefont {Zhang}, \citenamefont {Schulthess},\ and\ \citenamefont
  {MacLaren}}]{Butler2001_PhysRevB_63_054416}%
  \BibitemOpen
  \bibfield  {author} {\bibinfo {author} {\bibfnamefont {W.~H.}\ \bibnamefont
  {Butler}}, \bibinfo {author} {\bibfnamefont {X.-G.}\ \bibnamefont {Zhang}},
  \bibinfo {author} {\bibfnamefont {T.~C.}\ \bibnamefont {Schulthess}},\ and\
  \bibinfo {author} {\bibfnamefont {J.~M.}\ \bibnamefont {MacLaren}},\
  }\bibfield  {title} {\bibinfo {title} {{Spin-dependent tunneling conductance
  of $\mathrm{Fe}|\mathrm{MgO}|\mathrm{Fe}$ sandwiches}},\ }\href
  {https://doi.org/10.1103/PhysRevB.63.054416} {\bibfield  {journal} {\bibinfo
  {journal} {Phys. Rev. B}\ }\textbf {\bibinfo {volume} {63}},\ \bibinfo
  {pages} {054416} (\bibinfo {year} {2001})}\BibitemShut {NoStop}%
\bibitem [{\citenamefont {Mathon}\ and\ \citenamefont
  {Umerski}(2001)}]{Mathon2001_PhysRevB_63_220403}%
  \BibitemOpen
  \bibfield  {author} {\bibinfo {author} {\bibfnamefont {J.}~\bibnamefont
  {Mathon}}\ and\ \bibinfo {author} {\bibfnamefont {A.}~\bibnamefont
  {Umerski}},\ }\bibfield  {title} {\bibinfo {title} {{Theory of tunneling
  magnetoresistance of an epitaxial Fe/MgO/Fe(001) junction}},\ }\href
  {https://doi.org/10.1103/PhysRevB.63.220403} {\bibfield  {journal} {\bibinfo
  {journal} {Phys. Rev. B}\ }\textbf {\bibinfo {volume} {63}},\ \bibinfo
  {pages} {220403} (\bibinfo {year} {2001})}\BibitemShut {NoStop}%
\bibitem [{\citenamefont {Parkin}\ \emph {et~al.}(2004)\citenamefont {Parkin},
  \citenamefont {Kaiser}, \citenamefont {Panchula}, \citenamefont {Rice},
  \citenamefont {Hughes}, \citenamefont {Samant},\ and\ \citenamefont
  {Yang}}]{Parkin2004_NatMater_3_862}%
  \BibitemOpen
  \bibfield  {author} {\bibinfo {author} {\bibfnamefont {S.~S.}\ \bibnamefont
  {Parkin}}, \bibinfo {author} {\bibfnamefont {C.}~\bibnamefont {Kaiser}},
  \bibinfo {author} {\bibfnamefont {A.}~\bibnamefont {Panchula}}, \bibinfo
  {author} {\bibfnamefont {P.~M.}\ \bibnamefont {Rice}}, \bibinfo {author}
  {\bibfnamefont {B.}~\bibnamefont {Hughes}}, \bibinfo {author} {\bibfnamefont
  {M.}~\bibnamefont {Samant}},\ and\ \bibinfo {author} {\bibfnamefont {S.-H.}\
  \bibnamefont {Yang}},\ }\bibfield  {title} {\bibinfo {title} {{Giant
  tunnelling magnetoresistance at room temperature with MgO (100) tunnel
  barriers}},\ }\href@noop {} {\bibfield  {journal} {\bibinfo  {journal} {Nat.
  Mater.}\ }\textbf {\bibinfo {volume} {3}},\ \bibinfo {pages} {862--867}
  (\bibinfo {year} {2004})}\BibitemShut {NoStop}%
\bibitem [{\citenamefont {Yuasa}\ and\ \citenamefont
  {Djayaprawira}(2007)}]{Yuasa2007_JPhysD_40_R337}%
  \BibitemOpen
  \bibfield  {author} {\bibinfo {author} {\bibfnamefont {S.}~\bibnamefont
  {Yuasa}}\ and\ \bibinfo {author} {\bibfnamefont {D.~D.}\ \bibnamefont
  {Djayaprawira}},\ }\bibfield  {title} {\bibinfo {title} {{Giant tunnel
  magnetoresistance in magnetic tunnel junctions with a crystalline
  {MgO}(0{\hspace{0.167em}}0{\hspace{0.167em}}1) barrier}},\ }\href
  {https://doi.org/10.1088/0022-3727/40/21/r01} {\bibfield  {journal} {\bibinfo
   {journal} {J. Phys. D: Appl. Phys.}\ }\textbf {\bibinfo {volume} {40}},\
  \bibinfo {pages} {R337--R354} (\bibinfo {year} {2007})}\BibitemShut {NoStop}%
\bibitem [{\citenamefont {\ifmmode~\check{Z}\else \v{Z}\fi{}elezn\'y}\ \emph
  {et~al.}(2017)\citenamefont {\ifmmode~\check{Z}\else \v{Z}\fi{}elezn\'y},
  \citenamefont {Zhang}, \citenamefont {Felser},\ and\ \citenamefont
  {Yan}}]{Zelezny2017_PhysRevLett_119_187204}%
  \BibitemOpen
  \bibfield  {author} {\bibinfo {author} {\bibfnamefont {J.}~\bibnamefont
  {\ifmmode~\check{Z}\else \v{Z}\fi{}elezn\'y}}, \bibinfo {author}
  {\bibfnamefont {Y.}~\bibnamefont {Zhang}}, \bibinfo {author} {\bibfnamefont
  {C.}~\bibnamefont {Felser}},\ and\ \bibinfo {author} {\bibfnamefont
  {B.}~\bibnamefont {Yan}},\ }\bibfield  {title} {\bibinfo {title}
  {{Spin-Polarized Current in Noncollinear Antiferromagnets}},\ }\href
  {https://doi.org/10.1103/PhysRevLett.119.187204} {\bibfield  {journal}
  {\bibinfo  {journal} {Phys. Rev. Lett.}\ }\textbf {\bibinfo {volume} {119}},\
  \bibinfo {pages} {187204} (\bibinfo {year} {2017})}\BibitemShut {NoStop}%
\bibitem [{\citenamefont {Zhang}\ \emph {et~al.}(2018)\citenamefont {Zhang},
  \citenamefont {\v{Z}elezn\'{y}}, \citenamefont {Sun}, \citenamefont {van~den
  Brink},\ and\ \citenamefont {Yan}}]{Zhang2018_NewJPhys_20_073028}%
  \BibitemOpen
  \bibfield  {author} {\bibinfo {author} {\bibfnamefont {Y.}~\bibnamefont
  {Zhang}}, \bibinfo {author} {\bibfnamefont {J.}~\bibnamefont
  {\v{Z}elezn\'{y}}}, \bibinfo {author} {\bibfnamefont {Y.}~\bibnamefont
  {Sun}}, \bibinfo {author} {\bibfnamefont {J.}~\bibnamefont {van~den Brink}},\
  and\ \bibinfo {author} {\bibfnamefont {B.}~\bibnamefont {Yan}},\ }\bibfield
  {title} {\bibinfo {title} {{Spin Hall effect emerging from a noncollinear
  magnetic lattice without spin–orbit coupling}},\ }\href
  {https://doi.org/10.1088/1367-2630/aad1eb} {\bibfield  {journal} {\bibinfo
  {journal} {New J. Phys.}\ }\textbf {\bibinfo {volume} {20}},\ \bibinfo
  {pages} {073028} (\bibinfo {year} {2018})}\BibitemShut {NoStop}%
\bibitem [{\citenamefont {Naka}\ \emph {et~al.}(2019)\citenamefont {Naka},
  \citenamefont {Hayami}, \citenamefont {Kusunose}, \citenamefont {Yanagi},
  \citenamefont {Motome},\ and\ \citenamefont
  {Seo}}]{Naka2019_NatCommun_10_4305}%
  \BibitemOpen
  \bibfield  {author} {\bibinfo {author} {\bibfnamefont {M.}~\bibnamefont
  {Naka}}, \bibinfo {author} {\bibfnamefont {S.}~\bibnamefont {Hayami}},
  \bibinfo {author} {\bibfnamefont {H.}~\bibnamefont {Kusunose}}, \bibinfo
  {author} {\bibfnamefont {Y.}~\bibnamefont {Yanagi}}, \bibinfo {author}
  {\bibfnamefont {Y.}~\bibnamefont {Motome}},\ and\ \bibinfo {author}
  {\bibfnamefont {H.}~\bibnamefont {Seo}},\ }\bibfield  {title} {\bibinfo
  {title} {{Spin current generation in organic antiferromagnets}},\ }\href@noop
  {} {\bibfield  {journal} {\bibinfo  {journal} {Nat. Commun.}\ }\textbf
  {\bibinfo {volume} {10}},\ \bibinfo {pages} {4305} (\bibinfo {year}
  {2019})}\BibitemShut {NoStop}%
\bibitem [{\citenamefont {Gonz\'alez-Hern\'andez}\ \emph
  {et~al.}(2021)\citenamefont {Gonz\'alez-Hern\'andez}, \citenamefont
  {\ifmmode~\check{S}\else \v{S}\fi{}mejkal}, \citenamefont {V\'yborn\'y},
  \citenamefont {Yahagi}, \citenamefont {Sinova}, \citenamefont {Jungwirth},\
  and\ \citenamefont {\ifmmode~\check{Z}\else
  \v{Z}\fi{}elezn\'y}}]{Gonzalez-Hernandez2021_PhysRevLett_126_127701}%
  \BibitemOpen
  \bibfield  {author} {\bibinfo {author} {\bibfnamefont {R.}~\bibnamefont
  {Gonz\'alez-Hern\'andez}}, \bibinfo {author} {\bibfnamefont {L.}~\bibnamefont
  {\ifmmode~\check{S}\else \v{S}\fi{}mejkal}}, \bibinfo {author} {\bibfnamefont
  {K.}~\bibnamefont {V\'yborn\'y}}, \bibinfo {author} {\bibfnamefont
  {Y.}~\bibnamefont {Yahagi}}, \bibinfo {author} {\bibfnamefont
  {J.}~\bibnamefont {Sinova}}, \bibinfo {author} {\bibfnamefont {T.~c.~v.}\
  \bibnamefont {Jungwirth}},\ and\ \bibinfo {author} {\bibfnamefont
  {J.}~\bibnamefont {\ifmmode~\check{Z}\else \v{Z}\fi{}elezn\'y}},\ }\bibfield
  {title} {\bibinfo {title} {{Efficient Electrical Spin Splitter Based on
  Nonrelativistic Collinear Antiferromagnetism}},\ }\href
  {https://doi.org/10.1103/PhysRevLett.126.127701} {\bibfield  {journal}
  {\bibinfo  {journal} {Phys. Rev. Lett.}\ }\textbf {\bibinfo {volume} {126}},\
  \bibinfo {pages} {127701} (\bibinfo {year} {2021})}\BibitemShut {NoStop}%
\bibitem [{\citenamefont {Watanabe}\ \emph {et~al.}(2024)\citenamefont
  {Watanabe}, \citenamefont {Shinohara}, \citenamefont {Nomoto}, \citenamefont
  {Togo},\ and\ \citenamefont {Arita}}]{Watanabe2024_PhysRevB_109_094438}%
  \BibitemOpen
  \bibfield  {author} {\bibinfo {author} {\bibfnamefont {H.}~\bibnamefont
  {Watanabe}}, \bibinfo {author} {\bibfnamefont {K.}~\bibnamefont {Shinohara}},
  \bibinfo {author} {\bibfnamefont {T.}~\bibnamefont {Nomoto}}, \bibinfo
  {author} {\bibfnamefont {A.}~\bibnamefont {Togo}},\ and\ \bibinfo {author}
  {\bibfnamefont {R.}~\bibnamefont {Arita}},\ }\bibfield  {title} {\bibinfo
  {title} {{Symmetry analysis with spin crystallographic groups: Disentangling
  effects free of spin-orbit coupling in emergent electromagnetism}},\ }\href
  {https://doi.org/10.1103/PhysRevB.109.094438} {\bibfield  {journal} {\bibinfo
   {journal} {Phys. Rev. B}\ }\textbf {\bibinfo {volume} {109}},\ \bibinfo
  {pages} {094438} (\bibinfo {year} {2024})}\BibitemShut {NoStop}%
\bibitem [{\citenamefont {Shao}\ \emph {et~al.}(2021)\citenamefont {Shao},
  \citenamefont {Zhang}, \citenamefont {Li}, \citenamefont {Eom},\ and\
  \citenamefont {Tsymbal}}]{Shao2021_NatCommun_12_7061}%
  \BibitemOpen
  \bibfield  {author} {\bibinfo {author} {\bibfnamefont {D.-F.}\ \bibnamefont
  {Shao}}, \bibinfo {author} {\bibfnamefont {S.-H.}\ \bibnamefont {Zhang}},
  \bibinfo {author} {\bibfnamefont {M.}~\bibnamefont {Li}}, \bibinfo {author}
  {\bibfnamefont {C.-B.}\ \bibnamefont {Eom}},\ and\ \bibinfo {author}
  {\bibfnamefont {E.~Y.}\ \bibnamefont {Tsymbal}},\ }\bibfield  {title}
  {\bibinfo {title} {{Spin-neutral currents for spintronics}},\ }\href@noop {}
  {\bibfield  {journal} {\bibinfo  {journal} {Nat. Commun.}\ }\textbf {\bibinfo
  {volume} {12}},\ \bibinfo {pages} {1--8} (\bibinfo {year}
  {2021})}\BibitemShut {NoStop}%
\bibitem [{\citenamefont {\ifmmode~\check{S}\else \v{S}\fi{}mejkal}\ \emph
  {et~al.}(2022{\natexlab{a}})\citenamefont {\ifmmode~\check{S}\else
  \v{S}\fi{}mejkal}, \citenamefont {Hellenes}, \citenamefont
  {Gonz\'alez-Hern\'andez}, \citenamefont {Sinova},\ and\ \citenamefont
  {Jungwirth}}]{Smejkal2022_PhysRevX_12_011028}%
  \BibitemOpen
  \bibfield  {author} {\bibinfo {author} {\bibfnamefont {L.}~\bibnamefont
  {\ifmmode~\check{S}\else \v{S}\fi{}mejkal}}, \bibinfo {author} {\bibfnamefont
  {A.~B.}\ \bibnamefont {Hellenes}}, \bibinfo {author} {\bibfnamefont
  {R.}~\bibnamefont {Gonz\'alez-Hern\'andez}}, \bibinfo {author} {\bibfnamefont
  {J.}~\bibnamefont {Sinova}},\ and\ \bibinfo {author} {\bibfnamefont
  {T.}~\bibnamefont {Jungwirth}},\ }\bibfield  {title} {\bibinfo {title}
  {{Giant and Tunneling Magnetoresistance in Unconventional Collinear
  Antiferromagnets with Nonrelativistic Spin-Momentum Coupling}},\ }\href
  {https://doi.org/10.1103/PhysRevX.12.011028} {\bibfield  {journal} {\bibinfo
  {journal} {Phys. Rev. X}\ }\textbf {\bibinfo {volume} {12}},\ \bibinfo
  {pages} {011028} (\bibinfo {year} {2022}{\natexlab{a}})}\BibitemShut
  {NoStop}%
\bibitem [{\citenamefont {Dong}\ \emph {et~al.}(2022)\citenamefont {Dong},
  \citenamefont {Li}, \citenamefont {Gurung}, \citenamefont {Zhu},
  \citenamefont {Zhang}, \citenamefont {Zheng}, \citenamefont {Tsymbal},\ and\
  \citenamefont {Zhang}}]{Dong2022_PhysRevLett_128_197201}%
  \BibitemOpen
  \bibfield  {author} {\bibinfo {author} {\bibfnamefont {J.}~\bibnamefont
  {Dong}}, \bibinfo {author} {\bibfnamefont {X.}~\bibnamefont {Li}}, \bibinfo
  {author} {\bibfnamefont {G.}~\bibnamefont {Gurung}}, \bibinfo {author}
  {\bibfnamefont {M.}~\bibnamefont {Zhu}}, \bibinfo {author} {\bibfnamefont
  {P.}~\bibnamefont {Zhang}}, \bibinfo {author} {\bibfnamefont
  {F.}~\bibnamefont {Zheng}}, \bibinfo {author} {\bibfnamefont {E.~Y.}\
  \bibnamefont {Tsymbal}},\ and\ \bibinfo {author} {\bibfnamefont
  {J.}~\bibnamefont {Zhang}},\ }\bibfield  {title} {\bibinfo {title}
  {{Tunneling Magnetoresistance in Noncollinear Antiferromagnetic Tunnel
  Junctions}},\ }\href {https://doi.org/10.1103/PhysRevLett.128.197201}
  {\bibfield  {journal} {\bibinfo  {journal} {Phys. Rev. Lett.}\ }\textbf
  {\bibinfo {volume} {128}},\ \bibinfo {pages} {197201} (\bibinfo {year}
  {2022})}\BibitemShut {NoStop}%
\bibitem [{\citenamefont {Chen}\ \emph {et~al.}(2023)\citenamefont {Chen},
  \citenamefont {Higo}, \citenamefont {Tanaka}, \citenamefont {Nomoto},
  \citenamefont {Tsai}, \citenamefont {Idzuchi}, \citenamefont {Shiga},
  \citenamefont {Sakamoto}, \citenamefont {Ando}, \citenamefont {Kosaki},
  \citenamefont {Matsuo}, \citenamefont {Nishio-Hamane}, \citenamefont {Arita},
  \citenamefont {Miwa},\ and\ \citenamefont
  {Nakatsuji}}]{Chen2023_Nature_613_490}%
  \BibitemOpen
  \bibfield  {author} {\bibinfo {author} {\bibfnamefont {X.}~\bibnamefont
  {Chen}}, \bibinfo {author} {\bibfnamefont {T.}~\bibnamefont {Higo}}, \bibinfo
  {author} {\bibfnamefont {K.}~\bibnamefont {Tanaka}}, \bibinfo {author}
  {\bibfnamefont {T.}~\bibnamefont {Nomoto}}, \bibinfo {author} {\bibfnamefont
  {H.}~\bibnamefont {Tsai}}, \bibinfo {author} {\bibfnamefont {H.}~\bibnamefont
  {Idzuchi}}, \bibinfo {author} {\bibfnamefont {M.}~\bibnamefont {Shiga}},
  \bibinfo {author} {\bibfnamefont {S.}~\bibnamefont {Sakamoto}}, \bibinfo
  {author} {\bibfnamefont {R.}~\bibnamefont {Ando}}, \bibinfo {author}
  {\bibfnamefont {H.}~\bibnamefont {Kosaki}}, \bibinfo {author} {\bibfnamefont
  {T.}~\bibnamefont {Matsuo}}, \bibinfo {author} {\bibfnamefont
  {D.}~\bibnamefont {Nishio-Hamane}}, \bibinfo {author} {\bibfnamefont
  {R.}~\bibnamefont {Arita}}, \bibinfo {author} {\bibfnamefont
  {S.}~\bibnamefont {Miwa}},\ and\ \bibinfo {author} {\bibfnamefont
  {S.}~\bibnamefont {Nakatsuji}},\ }\bibfield  {title} {\bibinfo {title}
  {{Octupole-driven magnetoresistance in an antiferromagnetic tunnel
  junction}},\ }\href@noop {} {\bibfield  {journal} {\bibinfo  {journal}
  {Nature}\ }\textbf {\bibinfo {volume} {613}},\ \bibinfo {pages} {490--495}
  (\bibinfo {year} {2023})}\BibitemShut {NoStop}%
\bibitem [{\citenamefont {Qin}\ \emph {et~al.}(2023)\citenamefont {Qin},
  \citenamefont {Yan}, \citenamefont {Wang}, \citenamefont {Chen},
  \citenamefont {Meng}, \citenamefont {Dong}, \citenamefont {Zhu},
  \citenamefont {Cai}, \citenamefont {Feng}, \citenamefont {Zhou},
  \citenamefont {Liu}, \citenamefont {Zhang}, \citenamefont {Zeng},
  \citenamefont {Zhang}, \citenamefont {Jiang},\ and\ \citenamefont
  {Liu}}]{Qin2023_Nature_613_485}%
  \BibitemOpen
  \bibfield  {author} {\bibinfo {author} {\bibfnamefont {P.}~\bibnamefont
  {Qin}}, \bibinfo {author} {\bibfnamefont {H.}~\bibnamefont {Yan}}, \bibinfo
  {author} {\bibfnamefont {X.}~\bibnamefont {Wang}}, \bibinfo {author}
  {\bibfnamefont {H.}~\bibnamefont {Chen}}, \bibinfo {author} {\bibfnamefont
  {Z.}~\bibnamefont {Meng}}, \bibinfo {author} {\bibfnamefont {J.}~\bibnamefont
  {Dong}}, \bibinfo {author} {\bibfnamefont {M.}~\bibnamefont {Zhu}}, \bibinfo
  {author} {\bibfnamefont {J.}~\bibnamefont {Cai}}, \bibinfo {author}
  {\bibfnamefont {Z.}~\bibnamefont {Feng}}, \bibinfo {author} {\bibfnamefont
  {X.}~\bibnamefont {Zhou}}, \bibinfo {author} {\bibfnamefont {L.}~\bibnamefont
  {Liu}}, \bibinfo {author} {\bibfnamefont {T.}~\bibnamefont {Zhang}}, \bibinfo
  {author} {\bibfnamefont {Z.}~\bibnamefont {Zeng}}, \bibinfo {author}
  {\bibfnamefont {J.}~\bibnamefont {Zhang}}, \bibinfo {author} {\bibfnamefont
  {C.}~\bibnamefont {Jiang}},\ and\ \bibinfo {author} {\bibfnamefont
  {Z.}~\bibnamefont {Liu}},\ }\bibfield  {title} {\bibinfo {title}
  {{Room-temperature magnetoresistance in an all-antiferromagnetic tunnel
  junction}},\ }\href@noop {} {\bibfield  {journal} {\bibinfo  {journal}
  {Nature}\ }\textbf {\bibinfo {volume} {613}},\ \bibinfo {pages} {485--489}
  (\bibinfo {year} {2023})}\BibitemShut {NoStop}%
\bibitem [{\citenamefont {Shao}\ \emph {et~al.}(2023)\citenamefont {Shao},
  \citenamefont {Jiang}, \citenamefont {Ding}, \citenamefont {Zhang},
  \citenamefont {Wang}, \citenamefont {Xiao}, \citenamefont {Gurung},
  \citenamefont {Lu}, \citenamefont {Sun},\ and\ \citenamefont
  {Tsymbal}}]{Shao2023_PhysRevLett_130_216702}%
  \BibitemOpen
  \bibfield  {author} {\bibinfo {author} {\bibfnamefont {D.-F.}\ \bibnamefont
  {Shao}}, \bibinfo {author} {\bibfnamefont {Y.-Y.}\ \bibnamefont {Jiang}},
  \bibinfo {author} {\bibfnamefont {J.}~\bibnamefont {Ding}}, \bibinfo {author}
  {\bibfnamefont {S.-H.}\ \bibnamefont {Zhang}}, \bibinfo {author}
  {\bibfnamefont {Z.-A.}\ \bibnamefont {Wang}}, \bibinfo {author}
  {\bibfnamefont {R.-C.}\ \bibnamefont {Xiao}}, \bibinfo {author}
  {\bibfnamefont {G.}~\bibnamefont {Gurung}}, \bibinfo {author} {\bibfnamefont
  {W.~J.}\ \bibnamefont {Lu}}, \bibinfo {author} {\bibfnamefont {Y.~P.}\
  \bibnamefont {Sun}},\ and\ \bibinfo {author} {\bibfnamefont {E.~Y.}\
  \bibnamefont {Tsymbal}},\ }\bibfield  {title} {\bibinfo {title} {{N\'eel Spin
  Currents in Antiferromagnets}},\ }\href
  {https://doi.org/10.1103/PhysRevLett.130.216702} {\bibfield  {journal}
  {\bibinfo  {journal} {Phys. Rev. Lett.}\ }\textbf {\bibinfo {volume} {130}},\
  \bibinfo {pages} {216702} (\bibinfo {year} {2023})}\BibitemShut {NoStop}%
\bibitem [{\citenamefont {Cui}\ \emph {et~al.}(2023)\citenamefont {Cui},
  \citenamefont {Zhu}, \citenamefont {Yao}, \citenamefont {Cui},\ and\
  \citenamefont {Yang}}]{Cui2023_PhysRevB_108_024410}%
  \BibitemOpen
  \bibfield  {author} {\bibinfo {author} {\bibfnamefont {Q.}~\bibnamefont
  {Cui}}, \bibinfo {author} {\bibfnamefont {Y.}~\bibnamefont {Zhu}}, \bibinfo
  {author} {\bibfnamefont {X.}~\bibnamefont {Yao}}, \bibinfo {author}
  {\bibfnamefont {P.}~\bibnamefont {Cui}},\ and\ \bibinfo {author}
  {\bibfnamefont {H.}~\bibnamefont {Yang}},\ }\bibfield  {title} {\bibinfo
  {title} {{Giant spin-Hall and tunneling magnetoresistance effects based on a
  two-dimensional nonrelativistic antiferromagnetic metal}},\ }\href
  {https://doi.org/10.1103/PhysRevB.108.024410} {\bibfield  {journal} {\bibinfo
   {journal} {Phys. Rev. B}\ }\textbf {\bibinfo {volume} {108}},\ \bibinfo
  {pages} {024410} (\bibinfo {year} {2023})}\BibitemShut {NoStop}%
\bibitem [{\citenamefont {Jiang}\ \emph {et~al.}(2023)\citenamefont {Jiang},
  \citenamefont {Wang}, \citenamefont {Samanta}, \citenamefont {Zhang},
  \citenamefont {Xiao}, \citenamefont {Lu}, \citenamefont {Sun}, \citenamefont
  {Tsymbal},\ and\ \citenamefont {Shao}}]{Jiang2023_PhysRevB_108_174439}%
  \BibitemOpen
  \bibfield  {author} {\bibinfo {author} {\bibfnamefont {Y.-Y.}\ \bibnamefont
  {Jiang}}, \bibinfo {author} {\bibfnamefont {Z.-A.}\ \bibnamefont {Wang}},
  \bibinfo {author} {\bibfnamefont {K.}~\bibnamefont {Samanta}}, \bibinfo
  {author} {\bibfnamefont {S.-H.}\ \bibnamefont {Zhang}}, \bibinfo {author}
  {\bibfnamefont {R.-C.}\ \bibnamefont {Xiao}}, \bibinfo {author}
  {\bibfnamefont {W.~J.}\ \bibnamefont {Lu}}, \bibinfo {author} {\bibfnamefont
  {Y.~P.}\ \bibnamefont {Sun}}, \bibinfo {author} {\bibfnamefont {E.~Y.}\
  \bibnamefont {Tsymbal}},\ and\ \bibinfo {author} {\bibfnamefont {D.-F.}\
  \bibnamefont {Shao}},\ }\bibfield  {title} {\bibinfo {title} {{Prediction of
  giant tunneling magnetoresistance in
  $\mathrm{Ru}{\mathrm{O}}_{2}/\mathrm{Ti}{\mathrm{O}}_{2}/\mathrm{Ru}{\mathrm{O}}_{2}$
  (110) antiferromagnetic tunnel junctions}},\ }\href
  {https://doi.org/10.1103/PhysRevB.108.174439} {\bibfield  {journal} {\bibinfo
   {journal} {Phys. Rev. B}\ }\textbf {\bibinfo {volume} {108}},\ \bibinfo
  {pages} {174439} (\bibinfo {year} {2023})}\BibitemShut {NoStop}%
\bibitem [{\citenamefont {Chi}\ \emph {et~al.}(2024)\citenamefont {Chi},
  \citenamefont {Jiang}, \citenamefont {Zhu}, \citenamefont {Yu}, \citenamefont
  {Wan}, \citenamefont {Zhang},\ and\ \citenamefont
  {Han}}]{Chi2024_PhysRevApplied_21_034038}%
  \BibitemOpen
  \bibfield  {author} {\bibinfo {author} {\bibfnamefont {B.}~\bibnamefont
  {Chi}}, \bibinfo {author} {\bibfnamefont {L.}~\bibnamefont {Jiang}}, \bibinfo
  {author} {\bibfnamefont {Y.}~\bibnamefont {Zhu}}, \bibinfo {author}
  {\bibfnamefont {G.}~\bibnamefont {Yu}}, \bibinfo {author} {\bibfnamefont
  {C.}~\bibnamefont {Wan}}, \bibinfo {author} {\bibfnamefont {J.}~\bibnamefont
  {Zhang}},\ and\ \bibinfo {author} {\bibfnamefont {X.}~\bibnamefont {Han}},\
  }\bibfield  {title} {\bibinfo {title} {{Crystal-facet-oriented altermagnets
  for detecting ferromagnetic and antiferromagnetic states by giant tunneling
  magnetoresistance}},\ }\href
  {https://doi.org/10.1103/PhysRevApplied.21.034038} {\bibfield  {journal}
  {\bibinfo  {journal} {Phys. Rev. Appl.}\ }\textbf {\bibinfo {volume} {21}},\
  \bibinfo {pages} {034038} (\bibinfo {year} {2024})}\BibitemShut {NoStop}%
\bibitem [{\citenamefont {Samanta}\ \emph {et~al.}(2024)\citenamefont
  {Samanta}, \citenamefont {Jiang}, \citenamefont {Paudel}, \citenamefont
  {Shao},\ and\ \citenamefont {Tsymbal}}]{Samanta2024_PhysRevB_109_174407}%
  \BibitemOpen
  \bibfield  {author} {\bibinfo {author} {\bibfnamefont {K.}~\bibnamefont
  {Samanta}}, \bibinfo {author} {\bibfnamefont {Y.-Y.}\ \bibnamefont {Jiang}},
  \bibinfo {author} {\bibfnamefont {T.~R.}\ \bibnamefont {Paudel}}, \bibinfo
  {author} {\bibfnamefont {D.-F.}\ \bibnamefont {Shao}},\ and\ \bibinfo
  {author} {\bibfnamefont {E.~Y.}\ \bibnamefont {Tsymbal}},\ }\bibfield
  {title} {\bibinfo {title} {{Tunneling magnetoresistance in magnetic tunnel
  junctions with a single ferromagnetic electrode}},\ }\href
  {https://doi.org/10.1103/PhysRevB.109.174407} {\bibfield  {journal} {\bibinfo
   {journal} {Phys. Rev. B}\ }\textbf {\bibinfo {volume} {109}},\ \bibinfo
  {pages} {174407} (\bibinfo {year} {2024})}\BibitemShut {NoStop}%
\bibitem [{\citenamefont {Zhu}\ \emph {et~al.}(2024)\citenamefont {Zhu},
  \citenamefont {Dong}, \citenamefont {Li}, \citenamefont {Zheng},
  \citenamefont {Zhou}, \citenamefont {Wu},\ and\ \citenamefont
  {Zhang}}]{Zhu2024_ChinPhysLett_41_047502}%
  \BibitemOpen
  \bibfield  {author} {\bibinfo {author} {\bibfnamefont {M.}~\bibnamefont
  {Zhu}}, \bibinfo {author} {\bibfnamefont {J.}~\bibnamefont {Dong}}, \bibinfo
  {author} {\bibfnamefont {X.}~\bibnamefont {Li}}, \bibinfo {author}
  {\bibfnamefont {F.}~\bibnamefont {Zheng}}, \bibinfo {author} {\bibfnamefont
  {Y.}~\bibnamefont {Zhou}}, \bibinfo {author} {\bibfnamefont {K.}~\bibnamefont
  {Wu}},\ and\ \bibinfo {author} {\bibfnamefont {J.}~\bibnamefont {Zhang}},\
  }\bibfield  {title} {\bibinfo {title} {{Magnetic Switching Dynamics and
  Tunnel Magnetoresistance Effect Based on Spin-Splitting Noncollinear
  Antiferromagnet Mn$_{3}$Pt}},\ }\href
  {https://doi.org/10.1088/0256-307X/41/4/047502} {\bibfield  {journal}
  {\bibinfo  {journal} {Chin. Phys. Lett.}\ }\textbf {\bibinfo {volume} {41}},\
  \bibinfo {pages} {047502} (\bibinfo {year} {2024})}\BibitemShut {NoStop}%
\bibitem [{\citenamefont {Chou}\ \emph {et~al.}(2024)\citenamefont {Chou},
  \citenamefont {Ghosh}, \citenamefont {McGoldrick}, \citenamefont {Nguyen},
  \citenamefont {Gurung}, \citenamefont {Tsymbal}, \citenamefont {Li},
  \citenamefont {Mkhoyan},\ and\ \citenamefont
  {Liu}}]{Chou2024_NatCommun_15_7840}%
  \BibitemOpen
  \bibfield  {author} {\bibinfo {author} {\bibfnamefont {C.-T.}\ \bibnamefont
  {Chou}}, \bibinfo {author} {\bibfnamefont {S.}~\bibnamefont {Ghosh}},
  \bibinfo {author} {\bibfnamefont {B.~C.}\ \bibnamefont {McGoldrick}},
  \bibinfo {author} {\bibfnamefont {T.}~\bibnamefont {Nguyen}}, \bibinfo
  {author} {\bibfnamefont {G.}~\bibnamefont {Gurung}}, \bibinfo {author}
  {\bibfnamefont {E.~Y.}\ \bibnamefont {Tsymbal}}, \bibinfo {author}
  {\bibfnamefont {M.}~\bibnamefont {Li}}, \bibinfo {author} {\bibfnamefont
  {K.~A.}\ \bibnamefont {Mkhoyan}},\ and\ \bibinfo {author} {\bibfnamefont
  {L.}~\bibnamefont {Liu}},\ }\bibfield  {title} {\bibinfo {title} {{Large Spin
  Polarization from symmetry-breaking Antiferromagnets in Antiferromagnetic
  Tunnel Junctions}},\ }\href@noop {} {\bibfield  {journal} {\bibinfo
  {journal} {Nat.Commun.}\ }\textbf {\bibinfo {volume} {15}},\ \bibinfo {pages}
  {7840} (\bibinfo {year} {2024})}\BibitemShut {NoStop}%
\bibitem [{\citenamefont {Zhang}\ \emph {et~al.}(2024)\citenamefont {Zhang},
  \citenamefont {Yin}, \citenamefont {Zhu}, \citenamefont {Cheng},
  \citenamefont {Wen},\ and\ \citenamefont
  {He}}]{Zhang2024_PhysRevB_110_024428}%
  \BibitemOpen
  \bibfield  {author} {\bibinfo {author} {\bibfnamefont {X.}~\bibnamefont
  {Zhang}}, \bibinfo {author} {\bibfnamefont {L.}~\bibnamefont {Yin}}, \bibinfo
  {author} {\bibfnamefont {S.}~\bibnamefont {Zhu}}, \bibinfo {author}
  {\bibfnamefont {R.}~\bibnamefont {Cheng}}, \bibinfo {author} {\bibfnamefont
  {Y.}~\bibnamefont {Wen}},\ and\ \bibinfo {author} {\bibfnamefont
  {J.}~\bibnamefont {He}},\ }\bibfield  {title} {\bibinfo {title}
  {{Spin-correlation transport and multiple resistive states in multiferroic
  tunnel junctions}},\ }\href {https://doi.org/10.1103/PhysRevB.110.024428}
  {\bibfield  {journal} {\bibinfo  {journal} {Phys. Rev. B}\ }\textbf {\bibinfo
  {volume} {110}},\ \bibinfo {pages} {024428} (\bibinfo {year}
  {2024})}\BibitemShut {NoStop}%
\bibitem [{\citenamefont {Tanaka}\ \emph {et~al.}(2024)\citenamefont {Tanaka},
  \citenamefont {Nomoto},\ and\ \citenamefont
  {Arita}}]{Tanaka2024_PhysRevB_110_064433}%
  \BibitemOpen
  \bibfield  {author} {\bibinfo {author} {\bibfnamefont {K.}~\bibnamefont
  {Tanaka}}, \bibinfo {author} {\bibfnamefont {T.}~\bibnamefont {Nomoto}},\
  and\ \bibinfo {author} {\bibfnamefont {R.}~\bibnamefont {Arita}},\ }\bibfield
   {title} {\bibinfo {title} {{First-principles study of the tunnel
  magnetoresistance effect with Cr-doped ${\mathrm{RuO}}_{2}$ electrode}},\
  }\href {https://doi.org/10.1103/PhysRevB.110.064433} {\bibfield  {journal}
  {\bibinfo  {journal} {Phys. Rev. B}\ }\textbf {\bibinfo {volume} {110}},\
  \bibinfo {pages} {064433} (\bibinfo {year} {2024})}\BibitemShut {NoStop}%
\bibitem [{\citenamefont {Gurung}\ \emph {et~al.}(2024)\citenamefont {Gurung},
  \citenamefont {Elekhtiar}, \citenamefont {Luo}, \citenamefont {Shao},\ and\
  \citenamefont {Tsymbal}}]{Gurung2024_NatCommun_15_10242}%
  \BibitemOpen
  \bibfield  {author} {\bibinfo {author} {\bibfnamefont {G.}~\bibnamefont
  {Gurung}}, \bibinfo {author} {\bibfnamefont {M.}~\bibnamefont {Elekhtiar}},
  \bibinfo {author} {\bibfnamefont {Q.-Q.}\ \bibnamefont {Luo}}, \bibinfo
  {author} {\bibfnamefont {D.-F.}\ \bibnamefont {Shao}},\ and\ \bibinfo
  {author} {\bibfnamefont {E.~Y.}\ \bibnamefont {Tsymbal}},\ }\bibfield
  {title} {\bibinfo {title} {{Nearly perfect spin polarization of noncollinear
  antiferromagnets}},\ }\href@noop {} {\bibfield  {journal} {\bibinfo
  {journal} {Nat. Commun.}\ }\textbf {\bibinfo {volume} {15}},\ \bibinfo
  {pages} {10242} (\bibinfo {year} {2024})}\BibitemShut {NoStop}%
\bibitem [{\citenamefont {Luo}\ \emph {et~al.}(2025)\citenamefont {Luo},
  \citenamefont {Guo}, \citenamefont {Zhou}, \citenamefont {Gurung},
  \citenamefont {Xu}, \citenamefont {Lu}, \citenamefont {Sun}, \citenamefont
  {Tsymbal},\ and\ \citenamefont {Shao}}]{Luo2025_PhysRevB_111_144417}%
  \BibitemOpen
  \bibfield  {author} {\bibinfo {author} {\bibfnamefont {Q.-Q.}\ \bibnamefont
  {Luo}}, \bibinfo {author} {\bibfnamefont {X.-Y.}\ \bibnamefont {Guo}},
  \bibinfo {author} {\bibfnamefont {H.}~\bibnamefont {Zhou}}, \bibinfo {author}
  {\bibfnamefont {G.}~\bibnamefont {Gurung}}, \bibinfo {author} {\bibfnamefont
  {J.-M.}\ \bibnamefont {Xu}}, \bibinfo {author} {\bibfnamefont {W.-J.}\
  \bibnamefont {Lu}}, \bibinfo {author} {\bibfnamefont {Y.-P.}\ \bibnamefont
  {Sun}}, \bibinfo {author} {\bibfnamefont {E.~Y.}\ \bibnamefont {Tsymbal}},\
  and\ \bibinfo {author} {\bibfnamefont {D.-F.}\ \bibnamefont {Shao}},\
  }\bibfield  {title} {\bibinfo {title} {{Angular-dependent tunneling
  magnetoresistance in a tunnel junction with ferromagnetic and noncollinear
  antiferromagnetic electrodes}},\ }\href
  {https://doi.org/10.1103/PhysRevB.111.144417} {\bibfield  {journal} {\bibinfo
   {journal} {Phys. Rev. B}\ }\textbf {\bibinfo {volume} {111}},\ \bibinfo
  {pages} {144417} (\bibinfo {year} {2025})}\BibitemShut {NoStop}%
\bibitem [{\citenamefont {Wang}\ \emph {et~al.}(2024)\citenamefont {Wang},
  \citenamefont {Bian}, \citenamefont {Zhang},\ and\ \citenamefont
  {Yu}}]{Wang2024_ApplPhysLett_125_202404}%
  \BibitemOpen
  \bibfield  {author} {\bibinfo {author} {\bibfnamefont {Z.}~\bibnamefont
  {Wang}}, \bibinfo {author} {\bibfnamefont {B.}~\bibnamefont {Bian}}, \bibinfo
  {author} {\bibfnamefont {L.}~\bibnamefont {Zhang}},\ and\ \bibinfo {author}
  {\bibfnamefont {Z.}~\bibnamefont {Yu}},\ }\bibfield  {title} {\bibinfo
  {title} {{$\mathrm{Mn_{3}Sn}$-based noncollinear antiferromagnetic tunnel
  junctions with bilayer boron nitride tunnel barriers}},\ }\href
  {https://doi.org/10.1063/5.0234130} {\bibfield  {journal} {\bibinfo
  {journal} {Appl. Phys. Lett.}\ }\textbf {\bibinfo {volume} {125}},\ \bibinfo
  {pages} {202404} (\bibinfo {year} {2024})}\BibitemShut {NoStop}%
\bibitem [{\citenamefont {Yang}\ \emph
  {et~al.}(2025{\natexlab{a}})\citenamefont {Yang}, \citenamefont {Jiang},
  \citenamefont {Guo}, \citenamefont {Zhang}, \citenamefont {Xiao},
  \citenamefont {Lu}, \citenamefont {Wang}, \citenamefont {Sun}, \citenamefont
  {Tsymbal},\ and\ \citenamefont {Shao}}]{Yang2025_Newton_1_100142}%
  \BibitemOpen
  \bibfield  {author} {\bibinfo {author} {\bibfnamefont {L.}~\bibnamefont
  {Yang}}, \bibinfo {author} {\bibfnamefont {Y.-Y.}\ \bibnamefont {Jiang}},
  \bibinfo {author} {\bibfnamefont {X.-Y.}\ \bibnamefont {Guo}}, \bibinfo
  {author} {\bibfnamefont {S.-H.}\ \bibnamefont {Zhang}}, \bibinfo {author}
  {\bibfnamefont {R.-C.}\ \bibnamefont {Xiao}}, \bibinfo {author}
  {\bibfnamefont {W.-J.}\ \bibnamefont {Lu}}, \bibinfo {author} {\bibfnamefont
  {L.}~\bibnamefont {Wang}}, \bibinfo {author} {\bibfnamefont {Y.-P.}\
  \bibnamefont {Sun}}, \bibinfo {author} {\bibfnamefont {E.~Y.}\ \bibnamefont
  {Tsymbal}},\ and\ \bibinfo {author} {\bibfnamefont {D.-F.}\ \bibnamefont
  {Shao}},\ }\bibfield  {title} {\bibinfo {title} {{Interface-controlled
  antiferromagnetic tunnel junctions}},\ }\href
  {https://doi.org/https://doi.org/10.1016/j.newton.2025.100142} {\bibfield
  {journal} {\bibinfo  {journal} {Newton}\ ,\ \bibinfo {pages} {100142}}
  (\bibinfo {year} {2025}{\natexlab{a}})}\BibitemShut {NoStop}%
\bibitem [{\citenamefont {Liu}\ \emph {et~al.}(2025)\citenamefont {Liu},
  \citenamefont {Chen}, \citenamefont {Wu}, \citenamefont {Fan}, \citenamefont
  {Zhu}, \citenamefont {Bi}, \citenamefont {Liu}, \citenamefont {Shi},
  \citenamefont {Zhang}, \citenamefont {Wang}, \citenamefont {Li},
  \citenamefont {Yang}, \citenamefont {Lu}, \citenamefont {Zhou},\ and\
  \citenamefont {Liu}}]{Liu2025_AdvSci_12_e02985}%
  \BibitemOpen
  \bibfield  {author} {\bibinfo {author} {\bibfnamefont {S.}~\bibnamefont
  {Liu}}, \bibinfo {author} {\bibfnamefont {T.}~\bibnamefont {Chen}}, \bibinfo
  {author} {\bibfnamefont {B.}~\bibnamefont {Wu}}, \bibinfo {author}
  {\bibfnamefont {H.}~\bibnamefont {Fan}}, \bibinfo {author} {\bibfnamefont
  {Y.}~\bibnamefont {Zhu}}, \bibinfo {author} {\bibfnamefont {S.}~\bibnamefont
  {Bi}}, \bibinfo {author} {\bibfnamefont {Y.}~\bibnamefont {Liu}}, \bibinfo
  {author} {\bibfnamefont {Y.}~\bibnamefont {Shi}}, \bibinfo {author}
  {\bibfnamefont {W.}~\bibnamefont {Zhang}}, \bibinfo {author} {\bibfnamefont
  {M.}~\bibnamefont {Wang}}, \bibinfo {author} {\bibfnamefont {Q.}~\bibnamefont
  {Li}}, \bibinfo {author} {\bibfnamefont {J.}~\bibnamefont {Yang}}, \bibinfo
  {author} {\bibfnamefont {J.}~\bibnamefont {Lu}}, \bibinfo {author}
  {\bibfnamefont {T.}~\bibnamefont {Zhou}},\ and\ \bibinfo {author}
  {\bibfnamefont {B.}~\bibnamefont {Liu}},\ }\bibfield  {title} {\bibinfo
  {title} {{$\mathrm{Mn_{3}SnN}$-Based Antiferromagnetic Tunnel Junction with
  Giant Tunneling Magnetoresistance and Multi-States: Design and Theoretical
  Validation}},\ }\href
  {https://doi.org/https://doi.org/10.1002/advs.202502985} {\bibfield
  {journal} {\bibinfo  {journal} {Adv. Sci.}\ }\textbf {\bibinfo {volume}
  {12}},\ \bibinfo {pages} {e02985} (\bibinfo {year} {2025})}\BibitemShut
  {NoStop}%
\bibitem [{\citenamefont {Noh}\ \emph {et~al.}(2025)\citenamefont {Noh},
  \citenamefont {Kim}, \citenamefont {Lee}, \citenamefont {Jung}, \citenamefont
  {Seo}, \citenamefont {So}, \citenamefont {Lee}, \citenamefont {Lee},
  \citenamefont {Park}, \citenamefont {Yang}, \citenamefont {Oh}, \citenamefont
  {Jin}, \citenamefont {Sohn},\ and\ \citenamefont
  {Yoo}}]{Noh2025_PhysRevLett_134_246703}%
  \BibitemOpen
  \bibfield  {author} {\bibinfo {author} {\bibfnamefont {S.}~\bibnamefont
  {Noh}}, \bibinfo {author} {\bibfnamefont {G.-H.}\ \bibnamefont {Kim}},
  \bibinfo {author} {\bibfnamefont {J.}~\bibnamefont {Lee}}, \bibinfo {author}
  {\bibfnamefont {H.}~\bibnamefont {Jung}}, \bibinfo {author} {\bibfnamefont
  {U.}~\bibnamefont {Seo}}, \bibinfo {author} {\bibfnamefont {G.}~\bibnamefont
  {So}}, \bibinfo {author} {\bibfnamefont {J.}~\bibnamefont {Lee}}, \bibinfo
  {author} {\bibfnamefont {S.}~\bibnamefont {Lee}}, \bibinfo {author}
  {\bibfnamefont {M.}~\bibnamefont {Park}}, \bibinfo {author} {\bibfnamefont
  {S.}~\bibnamefont {Yang}}, \bibinfo {author} {\bibfnamefont {Y.~S.}\
  \bibnamefont {Oh}}, \bibinfo {author} {\bibfnamefont {H.}~\bibnamefont
  {Jin}}, \bibinfo {author} {\bibfnamefont {C.}~\bibnamefont {Sohn}},\ and\
  \bibinfo {author} {\bibfnamefont {J.-W.}\ \bibnamefont {Yoo}},\ }\bibfield
  {title} {\bibinfo {title} {{Tunneling Magnetoresistance in Altermagnetic
  ${\mathrm{RuO}}_{2}$-Based Magnetic Tunnel Junctions}},\ }\href
  {https://doi.org/10.1103/nrk5-5zrj} {\bibfield  {journal} {\bibinfo
  {journal} {Phys. Rev. Lett.}\ }\textbf {\bibinfo {volume} {134}},\ \bibinfo
  {pages} {246703} (\bibinfo {year} {2025})}\BibitemShut {NoStop}%
\bibitem [{\citenamefont {Zhu}\ \emph {et~al.}(2025)\citenamefont {Zhu},
  \citenamefont {Liu}, \citenamefont {Cui}, \citenamefont {Jiang},
  \citenamefont {Yang}, \citenamefont {Zhou},\ and\ \citenamefont
  {Liu}}]{Zhu2025_ApplPhysLett_127_082401}%
  \BibitemOpen
  \bibfield  {author} {\bibinfo {author} {\bibfnamefont {Y.}~\bibnamefont
  {Zhu}}, \bibinfo {author} {\bibfnamefont {S.}~\bibnamefont {Liu}}, \bibinfo
  {author} {\bibfnamefont {Q.}~\bibnamefont {Cui}}, \bibinfo {author}
  {\bibfnamefont {J.}~\bibnamefont {Jiang}}, \bibinfo {author} {\bibfnamefont
  {H.}~\bibnamefont {Yang}}, \bibinfo {author} {\bibfnamefont {T.}~\bibnamefont
  {Zhou}},\ and\ \bibinfo {author} {\bibfnamefont {B.}~\bibnamefont {Liu}},\
  }\bibfield  {title} {\bibinfo {title} {{Tunneling magnetoresistance in
  altermagnetic tunnel junctions with the half-metal electrode}},\ }\href
  {https://doi.org/10.1063/5.0283614} {\bibfield  {journal} {\bibinfo
  {journal} {Appl. Phys. Lett.}\ }\textbf {\bibinfo {volume} {127}},\ \bibinfo
  {pages} {082401} (\bibinfo {year} {2025})}\BibitemShut {NoStop}%
\bibitem [{\citenamefont {Sun}\ \emph {et~al.}(2025)\citenamefont {Sun},
  \citenamefont {Mao}, \citenamefont {Zhuang},\ and\ \citenamefont
  {Sun}}]{Sun2025_PhysRevB_112_094411}%
  \BibitemOpen
  \bibfield  {author} {\bibinfo {author} {\bibfnamefont {Y.-F.}\ \bibnamefont
  {Sun}}, \bibinfo {author} {\bibfnamefont {Y.}~\bibnamefont {Mao}}, \bibinfo
  {author} {\bibfnamefont {Y.-C.}\ \bibnamefont {Zhuang}},\ and\ \bibinfo
  {author} {\bibfnamefont {Q.-F.}\ \bibnamefont {Sun}},\ }\bibfield  {title}
  {\bibinfo {title} {{Tunneling magnetoresistance effect in altermagnets}},\
  }\href {https://doi.org/10.1103/t8b5-l859} {\bibfield  {journal} {\bibinfo
  {journal} {Phys. Rev. B}\ }\textbf {\bibinfo {volume} {112}},\ \bibinfo
  {pages} {094411} (\bibinfo {year} {2025})}\BibitemShut {NoStop}%
\bibitem [{\citenamefont {Yang}\ \emph
  {et~al.}(2025{\natexlab{b}})\citenamefont {Yang}, \citenamefont {Yang},
  \citenamefont {Wang}, \citenamefont {Li}, \citenamefont {Peng}, \citenamefont
  {Lee}, \citenamefont {Ang}, \citenamefont {Lu}, \citenamefont {Ang},\ and\
  \citenamefont {Fang}}]{Yang2025_PhysRevB_112_205202}%
  \BibitemOpen
  \bibfield  {author} {\bibinfo {author} {\bibfnamefont {Z.}~\bibnamefont
  {Yang}}, \bibinfo {author} {\bibfnamefont {X.}~\bibnamefont {Yang}}, \bibinfo
  {author} {\bibfnamefont {J.}~\bibnamefont {Wang}}, \bibinfo {author}
  {\bibfnamefont {Q.}~\bibnamefont {Li}}, \bibinfo {author} {\bibfnamefont
  {R.}~\bibnamefont {Peng}}, \bibinfo {author} {\bibfnamefont {C.~H.}\
  \bibnamefont {Lee}}, \bibinfo {author} {\bibfnamefont {L.~K.}\ \bibnamefont
  {Ang}}, \bibinfo {author} {\bibfnamefont {J.}~\bibnamefont {Lu}}, \bibinfo
  {author} {\bibfnamefont {Y.~S.}\ \bibnamefont {Ang}},\ and\ \bibinfo {author}
  {\bibfnamefont {S.}~\bibnamefont {Fang}},\ }\bibfield  {title} {\bibinfo
  {title} {{Unconventional thickness scaling of coherent tunnel
  magnetoresistance in altermagnets}},\ }\href
  {https://doi.org/10.1103/2thy-fzzj} {\bibfield  {journal} {\bibinfo
  {journal} {Phys. Rev. B}\ }\textbf {\bibinfo {volume} {112}},\ \bibinfo
  {pages} {205202} (\bibinfo {year} {2025}{\natexlab{b}})}\BibitemShut
  {NoStop}%
\bibitem [{\citenamefont {Luo}\ \emph {et~al.}(2026)\citenamefont {Luo},
  \citenamefont {Guo}, \citenamefont {Lu}, \citenamefont {Sun}, \citenamefont
  {Tsymbal}, \citenamefont {Li},\ and\ \citenamefont
  {Shao}}]{Luo2026_AdvFunctMater_36_e28671}%
  \BibitemOpen
  \bibfield  {author} {\bibinfo {author} {\bibfnamefont {Q.-Q.}\ \bibnamefont
  {Luo}}, \bibinfo {author} {\bibfnamefont {X.-Y.}\ \bibnamefont {Guo}},
  \bibinfo {author} {\bibfnamefont {W.-J.}\ \bibnamefont {Lu}}, \bibinfo
  {author} {\bibfnamefont {Y.-P.}\ \bibnamefont {Sun}}, \bibinfo {author}
  {\bibfnamefont {E.~Y.}\ \bibnamefont {Tsymbal}}, \bibinfo {author}
  {\bibfnamefont {M.}~\bibnamefont {Li}},\ and\ \bibinfo {author}
  {\bibfnamefont {D.-F.}\ \bibnamefont {Shao}},\ }\bibfield  {title} {\bibinfo
  {title} {{Multiferroic Tunnel Junctions With Noncollinear Antiferromagnetic
  Electrodes}},\ }\href
  {https://doi.org/https://doi.org/10.1002/adfm.202528671} {\bibfield
  {journal} {\bibinfo  {journal} {Adv. Funct. Mater.}\ }\textbf {\bibinfo
  {volume} {36}},\ \bibinfo {pages} {e28671} (\bibinfo {year}
  {2026})}\BibitemShut {NoStop}%
\bibitem [{\citenamefont {Tanaka}\ \emph {et~al.}(2026)\citenamefont {Tanaka},
  \citenamefont {Toga}, \citenamefont {Minami}, \citenamefont {Nakatsuji},
  \citenamefont {Nomoto}, \citenamefont {Koretsune},\ and\ \citenamefont
  {Arita}}]{Tanaka2026_PhysRevMater_10_044405}%
  \BibitemOpen
  \bibfield  {author} {\bibinfo {author} {\bibfnamefont {K.}~\bibnamefont
  {Tanaka}}, \bibinfo {author} {\bibfnamefont {Y.}~\bibnamefont {Toga}},
  \bibinfo {author} {\bibfnamefont {S.}~\bibnamefont {Minami}}, \bibinfo
  {author} {\bibfnamefont {S.}~\bibnamefont {Nakatsuji}}, \bibinfo {author}
  {\bibfnamefont {T.}~\bibnamefont {Nomoto}}, \bibinfo {author} {\bibfnamefont
  {T.}~\bibnamefont {Koretsune}},\ and\ \bibinfo {author} {\bibfnamefont
  {R.}~\bibnamefont {Arita}},\ }\bibfield  {title} {\bibinfo {title} {{Ab
  initio study of magnetoresistance effect in
  ${\text{Mn}}_{3}\text{Sn}/\text{MgO}/{\text{Mn}}_{3}\text{Sn}$
  antiferromagnetic tunnel junction}},\ }\href
  {https://doi.org/10.1103/xt7z-sf3x} {\bibfield  {journal} {\bibinfo
  {journal} {Phys. Rev. Mater.}\ }\textbf {\bibinfo {volume} {10}},\ \bibinfo
  {pages} {044405} (\bibinfo {year} {2026})}\BibitemShut {NoStop}%
\bibitem [{\citenamefont {Mao}\ \emph {et~al.}(2026)\citenamefont {Mao},
  \citenamefont {Liu}, \citenamefont {Li}, \citenamefont {Yang},\ and\
  \citenamefont {Yang}}]{Mao2026_PhysRevB_113_174409}%
  \BibitemOpen
  \bibfield  {author} {\bibinfo {author} {\bibfnamefont {C.}~\bibnamefont
  {Mao}}, \bibinfo {author} {\bibfnamefont {S.}~\bibnamefont {Liu}}, \bibinfo
  {author} {\bibfnamefont {S.}~\bibnamefont {Li}}, \bibinfo {author}
  {\bibfnamefont {J.}~\bibnamefont {Yang}},\ and\ \bibinfo {author}
  {\bibfnamefont {J.}~\bibnamefont {Yang}},\ }\bibfield  {title} {\bibinfo
  {title} {{Interfacial bond tailored altermagnetic tunnel junctions}},\ }\href
  {https://doi.org/10.1103/xcwv-bvkr} {\bibfield  {journal} {\bibinfo
  {journal} {Phys. Rev. B}\ }\textbf {\bibinfo {volume} {113}},\ \bibinfo
  {pages} {174409} (\bibinfo {year} {2026})}\BibitemShut {NoStop}%
\bibitem [{\citenamefont {Guo}\ \emph {et~al.}(2026)\citenamefont {Guo},
  \citenamefont {Mavani}, \citenamefont {Fang}, \citenamefont {Tang},
  \citenamefont {Li}, \citenamefont {Wu}, \citenamefont {Li}, \citenamefont
  {Tsymbal},\ and\ \citenamefont {Zhang}}]{Guo2026_ACSNano_20_18900}%
  \BibitemOpen
  \bibfield  {author} {\bibinfo {author} {\bibfnamefont {J.}~\bibnamefont
  {Guo}}, \bibinfo {author} {\bibfnamefont {H.}~\bibnamefont {Mavani}},
  \bibinfo {author} {\bibfnamefont {W.}~\bibnamefont {Fang}}, \bibinfo {author}
  {\bibfnamefont {J.}~\bibnamefont {Tang}}, \bibinfo {author} {\bibfnamefont
  {W.}~\bibnamefont {Li}}, \bibinfo {author} {\bibfnamefont {W.}~\bibnamefont
  {Wu}}, \bibinfo {author} {\bibfnamefont {H.}~\bibnamefont {Li}}, \bibinfo
  {author} {\bibfnamefont {E.~Y.}\ \bibnamefont {Tsymbal}},\ and\ \bibinfo
  {author} {\bibfnamefont {L.}~\bibnamefont {Zhang}},\ }\bibfield  {title}
  {\bibinfo {title} {{Termination-Preserved Ultrahigh Tunneling
  Magnetoresistance in Altermagnetic $\mathrm{KV_{2}Se_{2}O}$ }},\ }\href
  {https://doi.org/10.1021/acsnano.6c05425} {\bibfield  {journal} {\bibinfo
  {journal} {ACS Nano}\ }\textbf {\bibinfo {volume} {20}},\ \bibinfo {pages}
  {18900--18910} (\bibinfo {year} {2026})}\BibitemShut {NoStop}%
\bibitem [{\citenamefont {Shao}\ and\ \citenamefont
  {Tsymbal}(2024)}]{Shao2024_npjSpintronics_2_13}%
  \BibitemOpen
  \bibfield  {author} {\bibinfo {author} {\bibfnamefont {D.-F.}\ \bibnamefont
  {Shao}}\ and\ \bibinfo {author} {\bibfnamefont {E.~Y.}\ \bibnamefont
  {Tsymbal}},\ }\bibfield  {title} {\bibinfo {title} {{Antiferromagnetic tunnel
  junctions for spintronics}},\ }\href@noop {} {\bibfield  {journal} {\bibinfo
  {journal} {npj Spintronics}\ }\textbf {\bibinfo {volume} {2}},\ \bibinfo
  {pages} {13} (\bibinfo {year} {2024})}\BibitemShut {NoStop}%
\bibitem [{\citenamefont {Tanaka}\ \emph {et~al.}(2025)\citenamefont {Tanaka},
  \citenamefont {Nomoto},\ and\ \citenamefont
  {Arita}}]{Tanaka2025_JPhysCondensMatter_37_183003}%
  \BibitemOpen
  \bibfield  {author} {\bibinfo {author} {\bibfnamefont {K.}~\bibnamefont
  {Tanaka}}, \bibinfo {author} {\bibfnamefont {T.}~\bibnamefont {Nomoto}},\
  and\ \bibinfo {author} {\bibfnamefont {R.}~\bibnamefont {Arita}},\ }\bibfield
   {title} {\bibinfo {title} {{Approaches to tunnel magnetoresistance effect
  with antiferromagnets}},\ }\href {https://doi.org/10.1088/1361-648X/adc05e}
  {\bibfield  {journal} {\bibinfo  {journal} {J. Phys.: Condens. Matter}\
  }\textbf {\bibinfo {volume} {37}},\ \bibinfo {pages} {183003} (\bibinfo
  {year} {2025})}\BibitemShut {NoStop}%
\bibitem [{\citenamefont {Kang}\ \emph {et~al.}(2025)\citenamefont {Kang},
  \citenamefont {Hamdi}, \citenamefont {Cheung}, \citenamefont {Yuan},
  \citenamefont {Elekhtiar}, \citenamefont {Rogers}, \citenamefont {Meo},
  \citenamefont {Lim}, \citenamefont {Tey}, \citenamefont {D'Addario},
  \citenamefont {Konakanchi}, \citenamefont {Matt}, \citenamefont {Athas},
  \citenamefont {Arpaci}, \citenamefont {Wan}, \citenamefont {Mehta},
  \citenamefont {Upadhyaya}, \citenamefont {Carpentieri}, \citenamefont
  {Dravid}, \citenamefont {Hersam}, \citenamefont {Katine}, \citenamefont
  {Fuchs}, \citenamefont {Finocchio}, \citenamefont {Tsymbal}, \citenamefont
  {Rondinelli},\ and\ \citenamefont {Amiri}}]{Kang2025_arXiv_2509.03026}%
  \BibitemOpen
  \bibfield  {author} {\bibinfo {author} {\bibfnamefont {J.}~\bibnamefont
  {Kang}}, \bibinfo {author} {\bibfnamefont {M.}~\bibnamefont {Hamdi}},
  \bibinfo {author} {\bibfnamefont {S.~K.}\ \bibnamefont {Cheung}}, \bibinfo
  {author} {\bibfnamefont {L.-D.}\ \bibnamefont {Yuan}}, \bibinfo {author}
  {\bibfnamefont {M.}~\bibnamefont {Elekhtiar}}, \bibinfo {author}
  {\bibfnamefont {W.}~\bibnamefont {Rogers}}, \bibinfo {author} {\bibfnamefont
  {A.}~\bibnamefont {Meo}}, \bibinfo {author} {\bibfnamefont {P.~G.}\
  \bibnamefont {Lim}}, \bibinfo {author} {\bibfnamefont {M.~S.~N.}\
  \bibnamefont {Tey}}, \bibinfo {author} {\bibfnamefont {A.}~\bibnamefont
  {D'Addario}}, \bibinfo {author} {\bibfnamefont {S.~T.}\ \bibnamefont
  {Konakanchi}}, \bibinfo {author} {\bibfnamefont {E.}~\bibnamefont {Matt}},
  \bibinfo {author} {\bibfnamefont {J.}~\bibnamefont {Athas}}, \bibinfo
  {author} {\bibfnamefont {S.}~\bibnamefont {Arpaci}}, \bibinfo {author}
  {\bibfnamefont {L.}~\bibnamefont {Wan}}, \bibinfo {author} {\bibfnamefont
  {S.~C.}\ \bibnamefont {Mehta}}, \bibinfo {author} {\bibfnamefont
  {P.}~\bibnamefont {Upadhyaya}}, \bibinfo {author} {\bibfnamefont
  {M.}~\bibnamefont {Carpentieri}}, \bibinfo {author} {\bibfnamefont {V.~P.}\
  \bibnamefont {Dravid}}, \bibinfo {author} {\bibfnamefont {M.~C.}\
  \bibnamefont {Hersam}}, \bibinfo {author} {\bibfnamefont {J.~A.}\
  \bibnamefont {Katine}}, \bibinfo {author} {\bibfnamefont {G.~D.}\
  \bibnamefont {Fuchs}}, \bibinfo {author} {\bibfnamefont {G.}~\bibnamefont
  {Finocchio}}, \bibinfo {author} {\bibfnamefont {E.~Y.}\ \bibnamefont
  {Tsymbal}}, \bibinfo {author} {\bibfnamefont {J.~M.}\ \bibnamefont
  {Rondinelli}},\ and\ \bibinfo {author} {\bibfnamefont {P.~K.}\ \bibnamefont
  {Amiri}},\ }\href@noop {} {\bibinfo {title} {{Octupole-driven spin-transfer
  torque switching of all-antiferromagnetic tunnel junctions}}} (\bibinfo
  {year} {2025}),\ \Eprint {https://arxiv.org/abs/arXiv:2509.03026}
  {arXiv:2509.03026} \BibitemShut {NoStop}%
\bibitem [{\citenamefont {Elekhtiar}\ \emph {et~al.}(2026)\citenamefont
  {Elekhtiar}, \citenamefont {Shao},\ and\ \citenamefont
  {Tsymbal}}]{Elekhtiar2026_arXiv_2605.25369}%
  \BibitemOpen
  \bibfield  {author} {\bibinfo {author} {\bibfnamefont {M.}~\bibnamefont
  {Elekhtiar}}, \bibinfo {author} {\bibfnamefont {D.-F.}\ \bibnamefont
  {Shao}},\ and\ \bibinfo {author} {\bibfnamefont {E.~Y.}\ \bibnamefont
  {Tsymbal}},\ }\href@noop {} {\bibinfo {title} {{Effects of Band Symmetry on
  Spin-Dependent Transport in Noncollinear Antiferromagnetic Tunnel
  Junctions}}} (\bibinfo {year} {2026}),\ \Eprint
  {https://arxiv.org/abs/arXiv:2605.25369} {arXiv:2605.25369} \BibitemShut
  {NoStop}%
\bibitem [{\citenamefont {Hiraishi}\ \emph {et~al.}(2024)\citenamefont
  {Hiraishi}, \citenamefont {Okabe}, \citenamefont {Koda}, \citenamefont
  {Kadono}, \citenamefont {Muroi}, \citenamefont {Hirai},\ and\ \citenamefont
  {Hiroi}}]{Hiraishi2024_PhysRevLett_132_166702}%
  \BibitemOpen
  \bibfield  {author} {\bibinfo {author} {\bibfnamefont {M.}~\bibnamefont
  {Hiraishi}}, \bibinfo {author} {\bibfnamefont {H.}~\bibnamefont {Okabe}},
  \bibinfo {author} {\bibfnamefont {A.}~\bibnamefont {Koda}}, \bibinfo {author}
  {\bibfnamefont {R.}~\bibnamefont {Kadono}}, \bibinfo {author} {\bibfnamefont
  {T.}~\bibnamefont {Muroi}}, \bibinfo {author} {\bibfnamefont
  {D.}~\bibnamefont {Hirai}},\ and\ \bibinfo {author} {\bibfnamefont
  {Z.}~\bibnamefont {Hiroi}},\ }\bibfield  {title} {\bibinfo {title}
  {{Nonmagnetic Ground State in ${\mathrm{RuO}}_{2}$ Revealed by Muon Spin
  Rotation}},\ }\href {https://doi.org/10.1103/PhysRevLett.132.166702}
  {\bibfield  {journal} {\bibinfo  {journal} {Phys. Rev. Lett.}\ }\textbf
  {\bibinfo {volume} {132}},\ \bibinfo {pages} {166702} (\bibinfo {year}
  {2024})}\BibitemShut {NoStop}%
\bibitem [{\citenamefont {Smolyanyuk}\ \emph {et~al.}(2024)\citenamefont
  {Smolyanyuk}, \citenamefont {Mazin}, \citenamefont {Garcia-Gassull},\ and\
  \citenamefont {Valent\'{\i}}}]{Smolyanyuk2024_PhysRevB_109_134424}%
  \BibitemOpen
  \bibfield  {author} {\bibinfo {author} {\bibfnamefont {A.}~\bibnamefont
  {Smolyanyuk}}, \bibinfo {author} {\bibfnamefont {I.~I.}\ \bibnamefont
  {Mazin}}, \bibinfo {author} {\bibfnamefont {L.}~\bibnamefont
  {Garcia-Gassull}},\ and\ \bibinfo {author} {\bibfnamefont {R.}~\bibnamefont
  {Valent\'{\i}}},\ }\bibfield  {title} {\bibinfo {title} {{Fragility of the
  magnetic order in the prototypical altermagnet ${\mathrm{RuO}}_{2}$}},\
  }\href {https://doi.org/10.1103/PhysRevB.109.134424} {\bibfield  {journal}
  {\bibinfo  {journal} {Phys. Rev. B}\ }\textbf {\bibinfo {volume} {109}},\
  \bibinfo {pages} {134424} (\bibinfo {year} {2024})}\BibitemShut {NoStop}%
\bibitem [{\citenamefont {Ke{\ss}ler}\ \emph {et~al.}(2024)\citenamefont
  {Ke{\ss}ler}, \citenamefont {Garcia-Gassull}, \citenamefont {Suter},
  \citenamefont {Prokscha}, \citenamefont {Salman}, \citenamefont {Khalyavin},
  \citenamefont {Manuel}, \citenamefont {Orlandi}, \citenamefont {Mazin},
  \citenamefont {Valent{\'\i}},\ and\ \citenamefont
  {Moser}}]{Kessler2024_npjSpintronics_2_50}%
  \BibitemOpen
  \bibfield  {author} {\bibinfo {author} {\bibfnamefont {P.}~\bibnamefont
  {Ke{\ss}ler}}, \bibinfo {author} {\bibfnamefont {L.}~\bibnamefont
  {Garcia-Gassull}}, \bibinfo {author} {\bibfnamefont {A.}~\bibnamefont
  {Suter}}, \bibinfo {author} {\bibfnamefont {T.}~\bibnamefont {Prokscha}},
  \bibinfo {author} {\bibfnamefont {Z.}~\bibnamefont {Salman}}, \bibinfo
  {author} {\bibfnamefont {D.}~\bibnamefont {Khalyavin}}, \bibinfo {author}
  {\bibfnamefont {P.}~\bibnamefont {Manuel}}, \bibinfo {author} {\bibfnamefont
  {F.}~\bibnamefont {Orlandi}}, \bibinfo {author} {\bibfnamefont {I.~I.}\
  \bibnamefont {Mazin}}, \bibinfo {author} {\bibfnamefont {R.}~\bibnamefont
  {Valent{\'\i}}},\ and\ \bibinfo {author} {\bibfnamefont {S.}~\bibnamefont
  {Moser}},\ }\bibfield  {title} {\bibinfo {title} {{Absence of magnetic order
  in $\mathrm{RuO_{2}}$: insights from $\mu$ SR spectroscopy and neutron
  diffraction}},\ }\href@noop {} {\bibfield  {journal} {\bibinfo  {journal}
  {npj Spintronics}\ }\textbf {\bibinfo {volume} {2}},\ \bibinfo {pages} {50}
  (\bibinfo {year} {2024})}\BibitemShut {NoStop}%
\bibitem [{\citenamefont {Liu}\ \emph {et~al.}(2024)\citenamefont {Liu},
  \citenamefont {Zhan}, \citenamefont {Li}, \citenamefont {Liu}, \citenamefont
  {Cheng}, \citenamefont {Shi}, \citenamefont {Deng}, \citenamefont {Zhang},
  \citenamefont {Li}, \citenamefont {Ding}, \citenamefont {Jiang},
  \citenamefont {Ye}, \citenamefont {Liu}, \citenamefont {Jiang}, \citenamefont
  {Wang}, \citenamefont {Li}, \citenamefont {Xie}, \citenamefont {Wang},
  \citenamefont {Qiao}, \citenamefont {Wen}, \citenamefont {Sun},\ and\
  \citenamefont {Shen}}]{Liu2024_PhysRevLett_133_176401}%
  \BibitemOpen
  \bibfield  {author} {\bibinfo {author} {\bibfnamefont {J.}~\bibnamefont
  {Liu}}, \bibinfo {author} {\bibfnamefont {J.}~\bibnamefont {Zhan}}, \bibinfo
  {author} {\bibfnamefont {T.}~\bibnamefont {Li}}, \bibinfo {author}
  {\bibfnamefont {J.}~\bibnamefont {Liu}}, \bibinfo {author} {\bibfnamefont
  {S.}~\bibnamefont {Cheng}}, \bibinfo {author} {\bibfnamefont
  {Y.}~\bibnamefont {Shi}}, \bibinfo {author} {\bibfnamefont {L.}~\bibnamefont
  {Deng}}, \bibinfo {author} {\bibfnamefont {M.}~\bibnamefont {Zhang}},
  \bibinfo {author} {\bibfnamefont {C.}~\bibnamefont {Li}}, \bibinfo {author}
  {\bibfnamefont {J.}~\bibnamefont {Ding}}, \bibinfo {author} {\bibfnamefont
  {Q.}~\bibnamefont {Jiang}}, \bibinfo {author} {\bibfnamefont
  {M.}~\bibnamefont {Ye}}, \bibinfo {author} {\bibfnamefont {Z.}~\bibnamefont
  {Liu}}, \bibinfo {author} {\bibfnamefont {Z.}~\bibnamefont {Jiang}}, \bibinfo
  {author} {\bibfnamefont {S.}~\bibnamefont {Wang}}, \bibinfo {author}
  {\bibfnamefont {Q.}~\bibnamefont {Li}}, \bibinfo {author} {\bibfnamefont
  {Y.}~\bibnamefont {Xie}}, \bibinfo {author} {\bibfnamefont {Y.}~\bibnamefont
  {Wang}}, \bibinfo {author} {\bibfnamefont {S.}~\bibnamefont {Qiao}}, \bibinfo
  {author} {\bibfnamefont {J.}~\bibnamefont {Wen}}, \bibinfo {author}
  {\bibfnamefont {Y.}~\bibnamefont {Sun}},\ and\ \bibinfo {author}
  {\bibfnamefont {D.}~\bibnamefont {Shen}},\ }\bibfield  {title} {\bibinfo
  {title} {{Absence of Altermagnetic Spin Splitting Character in Rutile Oxide
  ${\mathrm{RuO}}_{2}$}},\ }\href
  {https://doi.org/10.1103/PhysRevLett.133.176401} {\bibfield  {journal}
  {\bibinfo  {journal} {Phys. Rev. Lett.}\ }\textbf {\bibinfo {volume} {133}},\
  \bibinfo {pages} {176401} (\bibinfo {year} {2024})}\BibitemShut {NoStop}%
\bibitem [{\citenamefont {Osumi}\ \emph {et~al.}(2026)\citenamefont {Osumi},
  \citenamefont {Yamauchi}, \citenamefont {Souma}, \citenamefont {Paul},
  \citenamefont {Honma}, \citenamefont {Nakayama}, \citenamefont {Ozawa},
  \citenamefont {Kitamura}, \citenamefont {Horiba}, \citenamefont
  {Kumigashira}, \citenamefont {Bigi}, \citenamefont {Bertran}, \citenamefont
  {Oguchi}, \citenamefont {Takahashi}, \citenamefont {Maeno},\ and\
  \citenamefont {Sato}}]{Osumi2026_PhysRevB_113_085116}%
  \BibitemOpen
  \bibfield  {author} {\bibinfo {author} {\bibfnamefont {T.}~\bibnamefont
  {Osumi}}, \bibinfo {author} {\bibfnamefont {K.}~\bibnamefont {Yamauchi}},
  \bibinfo {author} {\bibfnamefont {S.}~\bibnamefont {Souma}}, \bibinfo
  {author} {\bibfnamefont {S.}~\bibnamefont {Paul}}, \bibinfo {author}
  {\bibfnamefont {A.}~\bibnamefont {Honma}}, \bibinfo {author} {\bibfnamefont
  {K.}~\bibnamefont {Nakayama}}, \bibinfo {author} {\bibfnamefont
  {K.}~\bibnamefont {Ozawa}}, \bibinfo {author} {\bibfnamefont
  {M.}~\bibnamefont {Kitamura}}, \bibinfo {author} {\bibfnamefont
  {K.}~\bibnamefont {Horiba}}, \bibinfo {author} {\bibfnamefont
  {H.}~\bibnamefont {Kumigashira}}, \bibinfo {author} {\bibfnamefont
  {C.}~\bibnamefont {Bigi}}, \bibinfo {author} {\bibfnamefont {F.~m.~c.}\
  \bibnamefont {Bertran}}, \bibinfo {author} {\bibfnamefont {T.}~\bibnamefont
  {Oguchi}}, \bibinfo {author} {\bibfnamefont {T.}~\bibnamefont {Takahashi}},
  \bibinfo {author} {\bibfnamefont {Y.}~\bibnamefont {Maeno}},\ and\ \bibinfo
  {author} {\bibfnamefont {T.}~\bibnamefont {Sato}},\ }\bibfield  {title}
  {\bibinfo {title} {{Spin-degenerate bulk bands and topological surface states
  associated with Dirac nodal lines in ${\mathrm{RuO}}_{2}$}},\ }\href
  {https://doi.org/10.1103/wvs6-hqfv} {\bibfield  {journal} {\bibinfo
  {journal} {Phys. Rev. B}\ }\textbf {\bibinfo {volume} {113}},\ \bibinfo
  {pages} {085116} (\bibinfo {year} {2026})}\BibitemShut {NoStop}%
\bibitem [{\citenamefont {Berlijn}\ \emph {et~al.}(2017)\citenamefont
  {Berlijn}, \citenamefont {Snijders}, \citenamefont {Delaire}, \citenamefont
  {Zhou}, \citenamefont {Maier}, \citenamefont {Cao}, \citenamefont {Chi},
  \citenamefont {Matsuda}, \citenamefont {Wang}, \citenamefont {Koehler},
  \citenamefont {Kent},\ and\ \citenamefont
  {Weitering}}]{Berlijn2017_PhysRevLett_118_077201}%
  \BibitemOpen
  \bibfield  {author} {\bibinfo {author} {\bibfnamefont {T.}~\bibnamefont
  {Berlijn}}, \bibinfo {author} {\bibfnamefont {P.~C.}\ \bibnamefont
  {Snijders}}, \bibinfo {author} {\bibfnamefont {O.}~\bibnamefont {Delaire}},
  \bibinfo {author} {\bibfnamefont {H.-D.}\ \bibnamefont {Zhou}}, \bibinfo
  {author} {\bibfnamefont {T.~A.}\ \bibnamefont {Maier}}, \bibinfo {author}
  {\bibfnamefont {H.-B.}\ \bibnamefont {Cao}}, \bibinfo {author} {\bibfnamefont
  {S.-X.}\ \bibnamefont {Chi}}, \bibinfo {author} {\bibfnamefont
  {M.}~\bibnamefont {Matsuda}}, \bibinfo {author} {\bibfnamefont
  {Y.}~\bibnamefont {Wang}}, \bibinfo {author} {\bibfnamefont {M.~R.}\
  \bibnamefont {Koehler}}, \bibinfo {author} {\bibfnamefont {P.~R.~C.}\
  \bibnamefont {Kent}},\ and\ \bibinfo {author} {\bibfnamefont {H.~H.}\
  \bibnamefont {Weitering}},\ }\bibfield  {title} {\bibinfo {title} {{Itinerant
  Antiferromagnetism in ${\mathrm{RuO}}_{2}$}},\ }\href
  {https://doi.org/10.1103/PhysRevLett.118.077201} {\bibfield  {journal}
  {\bibinfo  {journal} {Phys. Rev. Lett.}\ }\textbf {\bibinfo {volume} {118}},\
  \bibinfo {pages} {077201} (\bibinfo {year} {2017})}\BibitemShut {NoStop}%
\bibitem [{\citenamefont {Zhu}\ \emph {et~al.}(2019)\citenamefont {Zhu},
  \citenamefont {Strempfer}, \citenamefont {Rao}, \citenamefont {Occhialini},
  \citenamefont {Pelliciari}, \citenamefont {Choi}, \citenamefont {Kawaguchi},
  \citenamefont {You}, \citenamefont {Mitchell}, \citenamefont {Shao-Horn},\
  and\ \citenamefont {Comin}}]{Zhu2019_PhysRevLett_122_017202}%
  \BibitemOpen
  \bibfield  {author} {\bibinfo {author} {\bibfnamefont {Z.~H.}\ \bibnamefont
  {Zhu}}, \bibinfo {author} {\bibfnamefont {J.}~\bibnamefont {Strempfer}},
  \bibinfo {author} {\bibfnamefont {R.~R.}\ \bibnamefont {Rao}}, \bibinfo
  {author} {\bibfnamefont {C.~A.}\ \bibnamefont {Occhialini}}, \bibinfo
  {author} {\bibfnamefont {J.}~\bibnamefont {Pelliciari}}, \bibinfo {author}
  {\bibfnamefont {Y.}~\bibnamefont {Choi}}, \bibinfo {author} {\bibfnamefont
  {T.}~\bibnamefont {Kawaguchi}}, \bibinfo {author} {\bibfnamefont
  {H.}~\bibnamefont {You}}, \bibinfo {author} {\bibfnamefont {J.~F.}\
  \bibnamefont {Mitchell}}, \bibinfo {author} {\bibfnamefont {Y.}~\bibnamefont
  {Shao-Horn}},\ and\ \bibinfo {author} {\bibfnamefont {R.}~\bibnamefont
  {Comin}},\ }\bibfield  {title} {\bibinfo {title} {{Anomalous
  Antiferromagnetism in Metallic ${\mathrm{RuO}}_{2}$ Determined by Resonant
  X-ray Scattering}},\ }\href {https://doi.org/10.1103/PhysRevLett.122.017202}
  {\bibfield  {journal} {\bibinfo  {journal} {Phys. Rev. Lett.}\ }\textbf
  {\bibinfo {volume} {122}},\ \bibinfo {pages} {017202} (\bibinfo {year}
  {2019})}\BibitemShut {NoStop}%
\bibitem [{\citenamefont {Ahn}\ \emph {et~al.}(2019)\citenamefont {Ahn},
  \citenamefont {Hariki}, \citenamefont {Lee},\ and\ \citenamefont
  {Kune\ifmmode~\check{s}\else \v{s}\fi{}}}]{Ahn2019_PhysRevB_99_184432}%
  \BibitemOpen
  \bibfield  {author} {\bibinfo {author} {\bibfnamefont {K.-H.}\ \bibnamefont
  {Ahn}}, \bibinfo {author} {\bibfnamefont {A.}~\bibnamefont {Hariki}},
  \bibinfo {author} {\bibfnamefont {K.-W.}\ \bibnamefont {Lee}},\ and\ \bibinfo
  {author} {\bibfnamefont {J.}~\bibnamefont {Kune\ifmmode~\check{s}\else
  \v{s}\fi{}}},\ }\bibfield  {title} {\bibinfo {title} {{Antiferromagnetism in
  ${\mathrm{RuO}}_{2}$ as $d$-wave Pomeranchuk instability}},\ }\href
  {https://doi.org/10.1103/PhysRevB.99.184432} {\bibfield  {journal} {\bibinfo
  {journal} {Phys. Rev. B}\ }\textbf {\bibinfo {volume} {99}},\ \bibinfo
  {pages} {184432} (\bibinfo {year} {2019})}\BibitemShut {NoStop}%
\bibitem [{\citenamefont {\v{S}mejkal}\ \emph {et~al.}(2020)\citenamefont
  {\v{S}mejkal}, \citenamefont {Gonz\'{a}lez-Hern\'{a}ndez}, \citenamefont
  {Jungwirth},\ and\ \citenamefont {Sinova}}]{Smejkal2020_SciAdv_6_eaaz8809}%
  \BibitemOpen
  \bibfield  {author} {\bibinfo {author} {\bibfnamefont {L.}~\bibnamefont
  {\v{S}mejkal}}, \bibinfo {author} {\bibfnamefont {R.}~\bibnamefont
  {Gonz\'{a}lez-Hern\'{a}ndez}}, \bibinfo {author} {\bibfnamefont
  {T.}~\bibnamefont {Jungwirth}},\ and\ \bibinfo {author} {\bibfnamefont
  {J.}~\bibnamefont {Sinova}},\ }\bibfield  {title} {\bibinfo {title} {{Crystal
  time-reversal symmetry breaking and spontaneous Hall effect in collinear
  antiferromagnets}},\ }\href {https://doi.org/10.1126/sciadv.aaz8809}
  {\bibfield  {journal} {\bibinfo  {journal} {Sci. Adv.}\ }\textbf {\bibinfo
  {volume} {6}},\ \bibinfo {pages} {eaaz8809} (\bibinfo {year}
  {2020})}\BibitemShut {NoStop}%
\bibitem [{\citenamefont {\ifmmode~\check{S}\else \v{S}\fi{}mejkal}\ \emph
  {et~al.}(2022{\natexlab{b}})\citenamefont {\ifmmode~\check{S}\else
  \v{S}\fi{}mejkal}, \citenamefont {Sinova},\ and\ \citenamefont
  {Jungwirth}}]{Smejkal2022_PhysRevX_12_031042}%
  \BibitemOpen
  \bibfield  {author} {\bibinfo {author} {\bibfnamefont {L.}~\bibnamefont
  {\ifmmode~\check{S}\else \v{S}\fi{}mejkal}}, \bibinfo {author} {\bibfnamefont
  {J.}~\bibnamefont {Sinova}},\ and\ \bibinfo {author} {\bibfnamefont
  {T.}~\bibnamefont {Jungwirth}},\ }\bibfield  {title} {\bibinfo {title}
  {{Beyond Conventional Ferromagnetism and Antiferromagnetism: A Phase with
  Nonrelativistic Spin and Crystal Rotation Symmetry}},\ }\href
  {https://doi.org/10.1103/PhysRevX.12.031042} {\bibfield  {journal} {\bibinfo
  {journal} {Phys. Rev. X}\ }\textbf {\bibinfo {volume} {12}},\ \bibinfo
  {pages} {031042} (\bibinfo {year} {2022}{\natexlab{b}})}\BibitemShut
  {NoStop}%
\bibitem [{\citenamefont {\ifmmode~\check{S}\else \v{S}\fi{}mejkal}\ \emph
  {et~al.}(2022{\natexlab{c}})\citenamefont {\ifmmode~\check{S}\else
  \v{S}\fi{}mejkal}, \citenamefont {Sinova},\ and\ \citenamefont
  {Jungwirth}}]{Smejkal2022_PhysRevX_12_040501}%
  \BibitemOpen
  \bibfield  {author} {\bibinfo {author} {\bibfnamefont {L.}~\bibnamefont
  {\ifmmode~\check{S}\else \v{S}\fi{}mejkal}}, \bibinfo {author} {\bibfnamefont
  {J.}~\bibnamefont {Sinova}},\ and\ \bibinfo {author} {\bibfnamefont
  {T.}~\bibnamefont {Jungwirth}},\ }\bibfield  {title} {\bibinfo {title}
  {{Emerging Research Landscape of Altermagnetism}},\ }\href
  {https://doi.org/10.1103/PhysRevX.12.040501} {\bibfield  {journal} {\bibinfo
  {journal} {Phys. Rev. X}\ }\textbf {\bibinfo {volume} {12}},\ \bibinfo
  {pages} {040501} (\bibinfo {year} {2022}{\natexlab{c}})}\BibitemShut
  {NoStop}%
\bibitem [{\citenamefont {Fedchenko}\ \emph {et~al.}(2024)\citenamefont
  {Fedchenko}, \citenamefont {Min\'{a}r}, \citenamefont {Akashdeep},
  \citenamefont {D’Souza}, \citenamefont {Vasilyev}, \citenamefont {Tkach},
  \citenamefont {Odenbreit}, \citenamefont {Nguyen}, \citenamefont
  {Kutnyakhov}, \citenamefont {Wind}, \citenamefont {Wenthaus}, \citenamefont
  {Scholz}, \citenamefont {Rossnagel}, \citenamefont {Hoesch}, \citenamefont
  {Aeschlimann}, \citenamefont {Stadtm\"{u}ller}, \citenamefont {Kl\"{a}ui},
  \citenamefont {Sch\"{o}nhense}, \citenamefont {Jungwirth}, \citenamefont
  {Hellenes}, \citenamefont {Jakob}, \citenamefont {\v{S}mejkal}, \citenamefont
  {Sinova},\ and\ \citenamefont {Elmers}}]{Fedchenko2024_SciAdv_10_eadj4883}%
  \BibitemOpen
  \bibfield  {author} {\bibinfo {author} {\bibfnamefont {O.}~\bibnamefont
  {Fedchenko}}, \bibinfo {author} {\bibfnamefont {J.}~\bibnamefont
  {Min\'{a}r}}, \bibinfo {author} {\bibfnamefont {A.}~\bibnamefont
  {Akashdeep}}, \bibinfo {author} {\bibfnamefont {S.~W.}\ \bibnamefont
  {D’Souza}}, \bibinfo {author} {\bibfnamefont {D.}~\bibnamefont {Vasilyev}},
  \bibinfo {author} {\bibfnamefont {O.}~\bibnamefont {Tkach}}, \bibinfo
  {author} {\bibfnamefont {L.}~\bibnamefont {Odenbreit}}, \bibinfo {author}
  {\bibfnamefont {Q.}~\bibnamefont {Nguyen}}, \bibinfo {author} {\bibfnamefont
  {D.}~\bibnamefont {Kutnyakhov}}, \bibinfo {author} {\bibfnamefont
  {N.}~\bibnamefont {Wind}}, \bibinfo {author} {\bibfnamefont {L.}~\bibnamefont
  {Wenthaus}}, \bibinfo {author} {\bibfnamefont {M.}~\bibnamefont {Scholz}},
  \bibinfo {author} {\bibfnamefont {K.}~\bibnamefont {Rossnagel}}, \bibinfo
  {author} {\bibfnamefont {M.}~\bibnamefont {Hoesch}}, \bibinfo {author}
  {\bibfnamefont {M.}~\bibnamefont {Aeschlimann}}, \bibinfo {author}
  {\bibfnamefont {B.}~\bibnamefont {Stadtm\"{u}ller}}, \bibinfo {author}
  {\bibfnamefont {M.}~\bibnamefont {Kl\"{a}ui}}, \bibinfo {author}
  {\bibfnamefont {G.}~\bibnamefont {Sch\"{o}nhense}}, \bibinfo {author}
  {\bibfnamefont {T.}~\bibnamefont {Jungwirth}}, \bibinfo {author}
  {\bibfnamefont {A.~B.}\ \bibnamefont {Hellenes}}, \bibinfo {author}
  {\bibfnamefont {G.}~\bibnamefont {Jakob}}, \bibinfo {author} {\bibfnamefont
  {L.}~\bibnamefont {\v{S}mejkal}}, \bibinfo {author} {\bibfnamefont
  {J.}~\bibnamefont {Sinova}},\ and\ \bibinfo {author} {\bibfnamefont {H.-J.}\
  \bibnamefont {Elmers}},\ }\bibfield  {title} {\bibinfo {title} {{Observation
  of time-reversal symmetry breaking in the band structure of altermagnetic
  RuO$_{2}$}},\ }\href {https://doi.org/10.1126/sciadv.adj4883} {\bibfield
  {journal} {\bibinfo  {journal} {Sci. Adv.}\ }\textbf {\bibinfo {volume}
  {10}},\ \bibinfo {pages} {eadj4883} (\bibinfo {year} {2024})}\BibitemShut
  {NoStop}%
\bibitem [{\citenamefont {Bai}\ \emph {et~al.}(2022)\citenamefont {Bai},
  \citenamefont {Han}, \citenamefont {Feng}, \citenamefont {Zhou},
  \citenamefont {Su}, \citenamefont {Wang}, \citenamefont {Liao}, \citenamefont
  {Zhu}, \citenamefont {Chen}, \citenamefont {Pan}, \citenamefont {Fan},\ and\
  \citenamefont {Song}}]{Bai2022_PhysRevLett_128_197202}%
  \BibitemOpen
  \bibfield  {author} {\bibinfo {author} {\bibfnamefont {H.}~\bibnamefont
  {Bai}}, \bibinfo {author} {\bibfnamefont {L.}~\bibnamefont {Han}}, \bibinfo
  {author} {\bibfnamefont {X.~Y.}\ \bibnamefont {Feng}}, \bibinfo {author}
  {\bibfnamefont {Y.~J.}\ \bibnamefont {Zhou}}, \bibinfo {author}
  {\bibfnamefont {R.~X.}\ \bibnamefont {Su}}, \bibinfo {author} {\bibfnamefont
  {Q.}~\bibnamefont {Wang}}, \bibinfo {author} {\bibfnamefont {L.~Y.}\
  \bibnamefont {Liao}}, \bibinfo {author} {\bibfnamefont {W.~X.}\ \bibnamefont
  {Zhu}}, \bibinfo {author} {\bibfnamefont {X.~Z.}\ \bibnamefont {Chen}},
  \bibinfo {author} {\bibfnamefont {F.}~\bibnamefont {Pan}}, \bibinfo {author}
  {\bibfnamefont {X.~L.}\ \bibnamefont {Fan}},\ and\ \bibinfo {author}
  {\bibfnamefont {C.}~\bibnamefont {Song}},\ }\bibfield  {title} {\bibinfo
  {title} {{Observation of Spin Splitting Torque in a Collinear Antiferromagnet
  ${\mathrm{RuO}}_{2}$}},\ }\href
  {https://doi.org/10.1103/PhysRevLett.128.197202} {\bibfield  {journal}
  {\bibinfo  {journal} {Phys. Rev. Lett.}\ }\textbf {\bibinfo {volume} {128}},\
  \bibinfo {pages} {197202} (\bibinfo {year} {2022})}\BibitemShut {NoStop}%
\bibitem [{\citenamefont {Karube}\ \emph {et~al.}(2022)\citenamefont {Karube},
  \citenamefont {Tanaka}, \citenamefont {Sugawara}, \citenamefont {Kadoguchi},
  \citenamefont {Kohda},\ and\ \citenamefont
  {Nitta}}]{Karube2022_PhysRevLett_129_137201}%
  \BibitemOpen
  \bibfield  {author} {\bibinfo {author} {\bibfnamefont {S.}~\bibnamefont
  {Karube}}, \bibinfo {author} {\bibfnamefont {T.}~\bibnamefont {Tanaka}},
  \bibinfo {author} {\bibfnamefont {D.}~\bibnamefont {Sugawara}}, \bibinfo
  {author} {\bibfnamefont {N.}~\bibnamefont {Kadoguchi}}, \bibinfo {author}
  {\bibfnamefont {M.}~\bibnamefont {Kohda}},\ and\ \bibinfo {author}
  {\bibfnamefont {J.}~\bibnamefont {Nitta}},\ }\bibfield  {title} {\bibinfo
  {title} {{Observation of Spin-Splitter Torque in Collinear Antiferromagnetic
  ${\mathrm{RuO}}_{2}$}},\ }\href
  {https://doi.org/10.1103/PhysRevLett.129.137201} {\bibfield  {journal}
  {\bibinfo  {journal} {Phys. Rev. Lett.}\ }\textbf {\bibinfo {volume} {129}},\
  \bibinfo {pages} {137201} (\bibinfo {year} {2022})}\BibitemShut {NoStop}%
\bibitem [{\citenamefont {Bai}\ \emph {et~al.}(2023)\citenamefont {Bai},
  \citenamefont {Zhang}, \citenamefont {Zhou}, \citenamefont {Chen},
  \citenamefont {Wan}, \citenamefont {Han}, \citenamefont {Zhu}, \citenamefont
  {Liang}, \citenamefont {Su}, \citenamefont {Han}, \citenamefont {Pan},\ and\
  \citenamefont {Song}}]{Bai2023_PhysRevLett_130_216701}%
  \BibitemOpen
  \bibfield  {author} {\bibinfo {author} {\bibfnamefont {H.}~\bibnamefont
  {Bai}}, \bibinfo {author} {\bibfnamefont {Y.~C.}\ \bibnamefont {Zhang}},
  \bibinfo {author} {\bibfnamefont {Y.~J.}\ \bibnamefont {Zhou}}, \bibinfo
  {author} {\bibfnamefont {P.}~\bibnamefont {Chen}}, \bibinfo {author}
  {\bibfnamefont {C.~H.}\ \bibnamefont {Wan}}, \bibinfo {author} {\bibfnamefont
  {L.}~\bibnamefont {Han}}, \bibinfo {author} {\bibfnamefont {W.~X.}\
  \bibnamefont {Zhu}}, \bibinfo {author} {\bibfnamefont {S.~X.}\ \bibnamefont
  {Liang}}, \bibinfo {author} {\bibfnamefont {Y.~C.}\ \bibnamefont {Su}},
  \bibinfo {author} {\bibfnamefont {X.~F.}\ \bibnamefont {Han}}, \bibinfo
  {author} {\bibfnamefont {F.}~\bibnamefont {Pan}},\ and\ \bibinfo {author}
  {\bibfnamefont {C.}~\bibnamefont {Song}},\ }\bibfield  {title} {\bibinfo
  {title} {{Efficient Spin-to-Charge Conversion via Altermagnetic Spin
  Splitting Effect in Antiferromagnet ${\mathrm{RuO}}_{2}$}},\ }\href
  {https://doi.org/10.1103/PhysRevLett.130.216701} {\bibfield  {journal}
  {\bibinfo  {journal} {Phys. Rev. Lett.}\ }\textbf {\bibinfo {volume} {130}},\
  \bibinfo {pages} {216701} (\bibinfo {year} {2023})}\BibitemShut {NoStop}%
\bibitem [{\citenamefont {Wang}\ \emph {et~al.}(2023)\citenamefont {Wang},
  \citenamefont {Tanaka}, \citenamefont {Sakai}, \citenamefont {Wang},
  \citenamefont {Deng}, \citenamefont {Lyu}, \citenamefont {Li}, \citenamefont
  {Tian}, \citenamefont {Shen}, \citenamefont {Ogawa}, \citenamefont
  {Kanazawa}, \citenamefont {Yu}, \citenamefont {Arita},\ and\ \citenamefont
  {Kagawa}}]{Wang2023_NatCommun_14_8240}%
  \BibitemOpen
  \bibfield  {author} {\bibinfo {author} {\bibfnamefont {M.}~\bibnamefont
  {Wang}}, \bibinfo {author} {\bibfnamefont {K.}~\bibnamefont {Tanaka}},
  \bibinfo {author} {\bibfnamefont {S.}~\bibnamefont {Sakai}}, \bibinfo
  {author} {\bibfnamefont {Z.}~\bibnamefont {Wang}}, \bibinfo {author}
  {\bibfnamefont {K.}~\bibnamefont {Deng}}, \bibinfo {author} {\bibfnamefont
  {Y.}~\bibnamefont {Lyu}}, \bibinfo {author} {\bibfnamefont {C.}~\bibnamefont
  {Li}}, \bibinfo {author} {\bibfnamefont {D.}~\bibnamefont {Tian}}, \bibinfo
  {author} {\bibfnamefont {S.}~\bibnamefont {Shen}}, \bibinfo {author}
  {\bibfnamefont {N.}~\bibnamefont {Ogawa}}, \bibinfo {author} {\bibfnamefont
  {N.}~\bibnamefont {Kanazawa}}, \bibinfo {author} {\bibfnamefont
  {P.}~\bibnamefont {Yu}}, \bibinfo {author} {\bibfnamefont {R.}~\bibnamefont
  {Arita}},\ and\ \bibinfo {author} {\bibfnamefont {F.}~\bibnamefont
  {Kagawa}},\ }\bibfield  {title} {\bibinfo {title} {{Emergent zero-field
  anomalous Hall effect in a reconstructed rutile antiferromagnetic metal}},\
  }\href@noop {} {\bibfield  {journal} {\bibinfo  {journal} {Nat. Commun.}\
  }\textbf {\bibinfo {volume} {14}},\ \bibinfo {pages} {8240} (\bibinfo {year}
  {2023})}\BibitemShut {NoStop}%
\bibitem [{\citenamefont {Smolyanyuk}\ \emph {et~al.}(2025)\citenamefont
  {Smolyanyuk}, \citenamefont {\ifmmode~\check{S}\else \v{S}\fi{}mejkal},\ and\
  \citenamefont {Mazin}}]{Smolyanyuk2025_PhysRevB_111_064406}%
  \BibitemOpen
  \bibfield  {author} {\bibinfo {author} {\bibfnamefont {A.}~\bibnamefont
  {Smolyanyuk}}, \bibinfo {author} {\bibfnamefont {L.}~\bibnamefont
  {\ifmmode~\check{S}\else \v{S}\fi{}mejkal}},\ and\ \bibinfo {author}
  {\bibfnamefont {I.~I.}\ \bibnamefont {Mazin}},\ }\bibfield  {title} {\bibinfo
  {title} {{Origin of the anomalous Hall effect in Cr-doped
  ${\mathrm{RuO}}_{2}$}},\ }\href {https://doi.org/10.1103/PhysRevB.111.064406}
  {\bibfield  {journal} {\bibinfo  {journal} {Phys. Rev. B}\ }\textbf {\bibinfo
  {volume} {111}},\ \bibinfo {pages} {064406} (\bibinfo {year}
  {2025})}\BibitemShut {NoStop}%
\bibitem [{\citenamefont {Hohenberg}\ and\ \citenamefont
  {Kohn}(1964)}]{Hohenberg1964_PhysRev_136_B864}%
  \BibitemOpen
  \bibfield  {author} {\bibinfo {author} {\bibfnamefont {P.}~\bibnamefont
  {Hohenberg}}\ and\ \bibinfo {author} {\bibfnamefont {W.}~\bibnamefont
  {Kohn}},\ }\bibfield  {title} {\bibinfo {title} {{Inhomogeneous Electron
  Gas}},\ }\href {https://doi.org/10.1103/PhysRev.136.B864} {\bibfield
  {journal} {\bibinfo  {journal} {Phys. Rev.}\ }\textbf {\bibinfo {volume}
  {136}},\ \bibinfo {pages} {B864--B871} (\bibinfo {year} {1964})}\BibitemShut
  {NoStop}%
\bibitem [{\citenamefont {Kohn}\ and\ \citenamefont
  {Sham}(1965)}]{Kohn1965_PhysRev_140_A1133}%
  \BibitemOpen
  \bibfield  {author} {\bibinfo {author} {\bibfnamefont {W.}~\bibnamefont
  {Kohn}}\ and\ \bibinfo {author} {\bibfnamefont {L.~J.}\ \bibnamefont
  {Sham}},\ }\bibfield  {title} {\bibinfo {title} {{Self-Consistent Equations
  Including Exchange and Correlation Effects}},\ }\href
  {https://doi.org/10.1103/PhysRev.140.A1133} {\bibfield  {journal} {\bibinfo
  {journal} {Phys. Rev.}\ }\textbf {\bibinfo {volume} {140}},\ \bibinfo {pages}
  {A1133--A1138} (\bibinfo {year} {1965})}\BibitemShut {NoStop}%
\bibitem [{\citenamefont {Giannozzi}\ \emph {et~al.}(2009)\citenamefont
  {Giannozzi}, \citenamefont {Baroni}, \citenamefont {Bonini}, \citenamefont
  {Calandra}, \citenamefont {Car}, \citenamefont {Cavazzoni}, \citenamefont
  {Ceresoli}, \citenamefont {Chiarotti}, \citenamefont {Cococcioni},
  \citenamefont {Dabo}, \citenamefont {Corso}, \citenamefont {de~Gironcoli},
  \citenamefont {Fabris}, \citenamefont {Fratesi}, \citenamefont {Gebauer},
  \citenamefont {Gerstmann}, \citenamefont {Gougoussis}, \citenamefont
  {Kokalj}, \citenamefont {Lazzeri}, \citenamefont {Martin-Samos},
  \citenamefont {Marzari}, \citenamefont {Mauri}, \citenamefont {Mazzarello},
  \citenamefont {Paolini}, \citenamefont {Pasquarello}, \citenamefont
  {Paulatto}, \citenamefont {Sbraccia}, \citenamefont {Scandolo}, \citenamefont
  {Sclauzero}, \citenamefont {Seitsonen}, \citenamefont {Smogunov},
  \citenamefont {Umari},\ and\ \citenamefont
  {Wentzcovitch}}]{Giannozzi2009_JPhysCondensMatter_21_395502}%
  \BibitemOpen
  \bibfield  {author} {\bibinfo {author} {\bibfnamefont {P.}~\bibnamefont
  {Giannozzi}}, \bibinfo {author} {\bibfnamefont {S.}~\bibnamefont {Baroni}},
  \bibinfo {author} {\bibfnamefont {N.}~\bibnamefont {Bonini}}, \bibinfo
  {author} {\bibfnamefont {M.}~\bibnamefont {Calandra}}, \bibinfo {author}
  {\bibfnamefont {R.}~\bibnamefont {Car}}, \bibinfo {author} {\bibfnamefont
  {C.}~\bibnamefont {Cavazzoni}}, \bibinfo {author} {\bibfnamefont
  {D.}~\bibnamefont {Ceresoli}}, \bibinfo {author} {\bibfnamefont {G.~L.}\
  \bibnamefont {Chiarotti}}, \bibinfo {author} {\bibfnamefont {M.}~\bibnamefont
  {Cococcioni}}, \bibinfo {author} {\bibfnamefont {I.}~\bibnamefont {Dabo}},
  \bibinfo {author} {\bibfnamefont {A.~D.}\ \bibnamefont {Corso}}, \bibinfo
  {author} {\bibfnamefont {S.}~\bibnamefont {de~Gironcoli}}, \bibinfo {author}
  {\bibfnamefont {S.}~\bibnamefont {Fabris}}, \bibinfo {author} {\bibfnamefont
  {G.}~\bibnamefont {Fratesi}}, \bibinfo {author} {\bibfnamefont
  {R.}~\bibnamefont {Gebauer}}, \bibinfo {author} {\bibfnamefont
  {U.}~\bibnamefont {Gerstmann}}, \bibinfo {author} {\bibfnamefont
  {C.}~\bibnamefont {Gougoussis}}, \bibinfo {author} {\bibfnamefont
  {A.}~\bibnamefont {Kokalj}}, \bibinfo {author} {\bibfnamefont
  {M.}~\bibnamefont {Lazzeri}}, \bibinfo {author} {\bibfnamefont
  {L.}~\bibnamefont {Martin-Samos}}, \bibinfo {author} {\bibfnamefont
  {N.}~\bibnamefont {Marzari}}, \bibinfo {author} {\bibfnamefont
  {F.}~\bibnamefont {Mauri}}, \bibinfo {author} {\bibfnamefont
  {R.}~\bibnamefont {Mazzarello}}, \bibinfo {author} {\bibfnamefont
  {S.}~\bibnamefont {Paolini}}, \bibinfo {author} {\bibfnamefont
  {A.}~\bibnamefont {Pasquarello}}, \bibinfo {author} {\bibfnamefont
  {L.}~\bibnamefont {Paulatto}}, \bibinfo {author} {\bibfnamefont
  {C.}~\bibnamefont {Sbraccia}}, \bibinfo {author} {\bibfnamefont
  {S.}~\bibnamefont {Scandolo}}, \bibinfo {author} {\bibfnamefont
  {G.}~\bibnamefont {Sclauzero}}, \bibinfo {author} {\bibfnamefont {A.~P.}\
  \bibnamefont {Seitsonen}}, \bibinfo {author} {\bibfnamefont {A.}~\bibnamefont
  {Smogunov}}, \bibinfo {author} {\bibfnamefont {P.}~\bibnamefont {Umari}},\
  and\ \bibinfo {author} {\bibfnamefont {R.~M.}\ \bibnamefont {Wentzcovitch}},\
  }\bibfield  {title} {\bibinfo {title} {{{QUANTUM} {ESPRESSO}: a modular and
  open-source software project for quantum simulations of materials}},\ }\href
  {https://doi.org/10.1088/0953-8984/21/39/395502} {\bibfield  {journal}
  {\bibinfo  {journal} {J. Phys.: Condens. Matter}\ }\textbf {\bibinfo {volume}
  {21}},\ \bibinfo {pages} {395502} (\bibinfo {year} {2009})}\BibitemShut
  {NoStop}%
\bibitem [{\citenamefont {Giannozzi}\ \emph {et~al.}(2017)\citenamefont
  {Giannozzi}, \citenamefont {Andreussi}, \citenamefont {Brumme}, \citenamefont
  {Bunau}, \citenamefont {Nardelli}, \citenamefont {Calandra}, \citenamefont
  {Car}, \citenamefont {Cavazzoni}, \citenamefont {Ceresoli}, \citenamefont
  {Cococcioni}, \citenamefont {Colonna}, \citenamefont {Carnimeo},
  \citenamefont {Corso}, \citenamefont {de~Gironcoli}, \citenamefont {Delugas},
  \citenamefont {DiStasio}, \citenamefont {Ferretti}, \citenamefont {Floris},
  \citenamefont {Fratesi}, \citenamefont {Fugallo}, \citenamefont {Gebauer},
  \citenamefont {Gerstmann}, \citenamefont {Giustino}, \citenamefont {Gorni},
  \citenamefont {Jia}, \citenamefont {Kawamura}, \citenamefont {Ko},
  \citenamefont {Kokalj}, \citenamefont {K\"{u}"{\c{c}}\"{u}kbenli},
  \citenamefont {Lazzeri}, \citenamefont {Marsili}, \citenamefont {Marzari},
  \citenamefont {Mauri}, \citenamefont {Nguyen}, \citenamefont {Nguyen},
  \citenamefont {de-la Roza}, \citenamefont {Paulatto}, \citenamefont
  {Ponc{\'{e}}}, \citenamefont {Rocca}, \citenamefont {Sabatini}, \citenamefont
  {Santra}, \citenamefont {Schlipf}, \citenamefont {Seitsonen}, \citenamefont
  {Smogunov}, \citenamefont {Timrov}, \citenamefont {Thonhauser}, \citenamefont
  {Umari}, \citenamefont {Vast}, \citenamefont {Wu},\ and\ \citenamefont
  {Baroni}}]{Giannozzi2017_JPhysCondensMatter_29_465901}%
  \BibitemOpen
  \bibfield  {author} {\bibinfo {author} {\bibfnamefont {P.}~\bibnamefont
  {Giannozzi}}, \bibinfo {author} {\bibfnamefont {O.}~\bibnamefont
  {Andreussi}}, \bibinfo {author} {\bibfnamefont {T.}~\bibnamefont {Brumme}},
  \bibinfo {author} {\bibfnamefont {O.}~\bibnamefont {Bunau}}, \bibinfo
  {author} {\bibfnamefont {M.~B.}\ \bibnamefont {Nardelli}}, \bibinfo {author}
  {\bibfnamefont {M.}~\bibnamefont {Calandra}}, \bibinfo {author}
  {\bibfnamefont {R.}~\bibnamefont {Car}}, \bibinfo {author} {\bibfnamefont
  {C.}~\bibnamefont {Cavazzoni}}, \bibinfo {author} {\bibfnamefont
  {D.}~\bibnamefont {Ceresoli}}, \bibinfo {author} {\bibfnamefont
  {M.}~\bibnamefont {Cococcioni}}, \bibinfo {author} {\bibfnamefont
  {N.}~\bibnamefont {Colonna}}, \bibinfo {author} {\bibfnamefont
  {I.}~\bibnamefont {Carnimeo}}, \bibinfo {author} {\bibfnamefont {A.~D.}\
  \bibnamefont {Corso}}, \bibinfo {author} {\bibfnamefont {S.}~\bibnamefont
  {de~Gironcoli}}, \bibinfo {author} {\bibfnamefont {P.}~\bibnamefont
  {Delugas}}, \bibinfo {author} {\bibfnamefont {R.~A.}\ \bibnamefont
  {DiStasio}}, \bibinfo {author} {\bibfnamefont {A.}~\bibnamefont {Ferretti}},
  \bibinfo {author} {\bibfnamefont {A.}~\bibnamefont {Floris}}, \bibinfo
  {author} {\bibfnamefont {G.}~\bibnamefont {Fratesi}}, \bibinfo {author}
  {\bibfnamefont {G.}~\bibnamefont {Fugallo}}, \bibinfo {author} {\bibfnamefont
  {R.}~\bibnamefont {Gebauer}}, \bibinfo {author} {\bibfnamefont
  {U.}~\bibnamefont {Gerstmann}}, \bibinfo {author} {\bibfnamefont
  {F.}~\bibnamefont {Giustino}}, \bibinfo {author} {\bibfnamefont
  {T.}~\bibnamefont {Gorni}}, \bibinfo {author} {\bibfnamefont
  {J.}~\bibnamefont {Jia}}, \bibinfo {author} {\bibfnamefont {M.}~\bibnamefont
  {Kawamura}}, \bibinfo {author} {\bibfnamefont {H.-Y.}\ \bibnamefont {Ko}},
  \bibinfo {author} {\bibfnamefont {A.}~\bibnamefont {Kokalj}}, \bibinfo
  {author} {\bibfnamefont {E.}~\bibnamefont {K\"{u}"{\c{c}}\"{u}kbenli}},
  \bibinfo {author} {\bibfnamefont {M.}~\bibnamefont {Lazzeri}}, \bibinfo
  {author} {\bibfnamefont {M.}~\bibnamefont {Marsili}}, \bibinfo {author}
  {\bibfnamefont {N.}~\bibnamefont {Marzari}}, \bibinfo {author} {\bibfnamefont
  {F.}~\bibnamefont {Mauri}}, \bibinfo {author} {\bibfnamefont {N.~L.}\
  \bibnamefont {Nguyen}}, \bibinfo {author} {\bibfnamefont {H.-V.}\
  \bibnamefont {Nguyen}}, \bibinfo {author} {\bibfnamefont {A.~O.}\
  \bibnamefont {de-la Roza}}, \bibinfo {author} {\bibfnamefont
  {L.}~\bibnamefont {Paulatto}}, \bibinfo {author} {\bibfnamefont
  {S.}~\bibnamefont {Ponc{\'{e}}}}, \bibinfo {author} {\bibfnamefont
  {D.}~\bibnamefont {Rocca}}, \bibinfo {author} {\bibfnamefont
  {R.}~\bibnamefont {Sabatini}}, \bibinfo {author} {\bibfnamefont
  {B.}~\bibnamefont {Santra}}, \bibinfo {author} {\bibfnamefont
  {M.}~\bibnamefont {Schlipf}}, \bibinfo {author} {\bibfnamefont {A.~P.}\
  \bibnamefont {Seitsonen}}, \bibinfo {author} {\bibfnamefont {A.}~\bibnamefont
  {Smogunov}}, \bibinfo {author} {\bibfnamefont {I.}~\bibnamefont {Timrov}},
  \bibinfo {author} {\bibfnamefont {T.}~\bibnamefont {Thonhauser}}, \bibinfo
  {author} {\bibfnamefont {P.}~\bibnamefont {Umari}}, \bibinfo {author}
  {\bibfnamefont {N.}~\bibnamefont {Vast}}, \bibinfo {author} {\bibfnamefont
  {X.}~\bibnamefont {Wu}},\ and\ \bibinfo {author} {\bibfnamefont
  {S.}~\bibnamefont {Baroni}},\ }\bibfield  {title} {\bibinfo {title}
  {{Advanced capabilities for materials modelling with Quantum {ESPRESSO}}},\
  }\href {https://doi.org/10.1088/1361-648x/aa8f79} {\bibfield  {journal}
  {\bibinfo  {journal} {J. Phys.: Condens. Matter}\ }\textbf {\bibinfo {volume}
  {29}},\ \bibinfo {pages} {465901} (\bibinfo {year} {2017})}\BibitemShut
  {NoStop}%
\bibitem [{\citenamefont {Perdew}\ \emph {et~al.}(1996)\citenamefont {Perdew},
  \citenamefont {Burke},\ and\ \citenamefont
  {Ernzerhof}}]{Perdew1996_PhysRevLett_77_3865}%
  \BibitemOpen
  \bibfield  {author} {\bibinfo {author} {\bibfnamefont {J.~P.}\ \bibnamefont
  {Perdew}}, \bibinfo {author} {\bibfnamefont {K.}~\bibnamefont {Burke}},\ and\
  \bibinfo {author} {\bibfnamefont {M.}~\bibnamefont {Ernzerhof}},\ }\bibfield
  {title} {\bibinfo {title} {{Generalized Gradient Approximation Made
  Simple}},\ }\href {https://doi.org/10.1103/PhysRevLett.77.3865} {\bibfield
  {journal} {\bibinfo  {journal} {Phys. Rev. Lett.}\ }\textbf {\bibinfo
  {volume} {77}},\ \bibinfo {pages} {3865--3868} (\bibinfo {year}
  {1996})}\BibitemShut {NoStop}%
\bibitem [{\citenamefont {{Dal
  Corso}}(2014)}]{DalCorso2014_ComptMaterSci_95_337}%
  \BibitemOpen
  \bibfield  {author} {\bibinfo {author} {\bibfnamefont {A.}~\bibnamefont {{Dal
  Corso}}},\ }\bibfield  {title} {\bibinfo {title} {{Pseudopotentials periodic
  table: From H to Pu}},\ }\href
  {https://doi.org/https://doi.org/10.1016/j.commatsci.2014.07.043} {\bibfield
  {journal} {\bibinfo  {journal} {Comput. Mater. Sci.}\ }\textbf {\bibinfo
  {volume} {95}},\ \bibinfo {pages} {337--350} (\bibinfo {year} {2014})},\
  \bibinfo {note} {https://dalcorso.github.io/pslibrary/}\BibitemShut {NoStop}%
\bibitem [{\citenamefont {Nordheim}(1931)}]{Nordheim1931_AnnPhys_401_607}%
  \BibitemOpen
  \bibfield  {author} {\bibinfo {author} {\bibfnamefont {L.}~\bibnamefont
  {Nordheim}},\ }\bibfield  {title} {\bibinfo {title} {{Zur Elektronentheorie
  der Metalle. I}},\ }\href
  {https://doi.org/https://doi.org/10.1002/andp.19314010507} {\bibfield
  {journal} {\bibinfo  {journal} {Ann. Phys.}\ }\textbf {\bibinfo {volume}
  {401}},\ \bibinfo {pages} {607--640} (\bibinfo {year} {1931})}\BibitemShut
  {NoStop}%
\bibitem [{\citenamefont {Bellaiche}\ and\ \citenamefont
  {Vanderbilt}(2000)}]{Bellaiche2000_PhysRevB_61_7877}%
  \BibitemOpen
  \bibfield  {author} {\bibinfo {author} {\bibfnamefont {L.}~\bibnamefont
  {Bellaiche}}\ and\ \bibinfo {author} {\bibfnamefont {D.}~\bibnamefont
  {Vanderbilt}},\ }\bibfield  {title} {\bibinfo {title} {{Virtual crystal
  approximation revisited: Application to dielectric and piezoelectric
  properties of perovskites}},\ }\href
  {https://doi.org/10.1103/PhysRevB.61.7877} {\bibfield  {journal} {\bibinfo
  {journal} {Phys. Rev. B}\ }\textbf {\bibinfo {volume} {61}},\ \bibinfo
  {pages} {7877--7882} (\bibinfo {year} {2000})}\BibitemShut {NoStop}%
\bibitem [{\citenamefont {Joon~Choi}\ and\ \citenamefont
  {Ihm}(1999)}]{Choi1999_PhysRevB_59_2267}%
  \BibitemOpen
  \bibfield  {author} {\bibinfo {author} {\bibfnamefont {H.}~\bibnamefont
  {Joon~Choi}}\ and\ \bibinfo {author} {\bibfnamefont {J.}~\bibnamefont
  {Ihm}},\ }\bibfield  {title} {\bibinfo {title} {{Ab initio pseudopotential
  method for the calculation of conductance in quantum wires}},\ }\href
  {https://doi.org/10.1103/PhysRevB.59.2267} {\bibfield  {journal} {\bibinfo
  {journal} {Phys. Rev. B}\ }\textbf {\bibinfo {volume} {59}},\ \bibinfo
  {pages} {2267--2275} (\bibinfo {year} {1999})}\BibitemShut {NoStop}%
\bibitem [{\citenamefont {Smogunov}\ \emph {et~al.}(2004)\citenamefont
  {Smogunov}, \citenamefont {Dal~Corso},\ and\ \citenamefont
  {Tosatti}}]{Smogunov2004_PhysRevB_70_045417}%
  \BibitemOpen
  \bibfield  {author} {\bibinfo {author} {\bibfnamefont {A.}~\bibnamefont
  {Smogunov}}, \bibinfo {author} {\bibfnamefont {A.}~\bibnamefont
  {Dal~Corso}},\ and\ \bibinfo {author} {\bibfnamefont {E.}~\bibnamefont
  {Tosatti}},\ }\bibfield  {title} {\bibinfo {title} {{Ballistic conductance of
  magnetic Co and Ni nanowires with ultrasoft pseudopotentials}},\ }\href
  {https://doi.org/10.1103/PhysRevB.70.045417} {\bibfield  {journal} {\bibinfo
  {journal} {Phys. Rev. B}\ }\textbf {\bibinfo {volume} {70}},\ \bibinfo
  {pages} {045417} (\bibinfo {year} {2004})}\BibitemShut {NoStop}%
\bibitem [{\citenamefont {Landauer}(1957)}]{Landauer1957_IBMJResDev_1_3}%
  \BibitemOpen
  \bibfield  {author} {\bibinfo {author} {\bibfnamefont {R.}~\bibnamefont
  {Landauer}},\ }\bibfield  {title} {\bibinfo {title} {{Spatial Variation of
  Currents and Fields Due to Localized Scatterers in Metallic Conduction}},\
  }\href {https://doi.org/10.1147/rd.13.0223} {\bibfield  {journal} {\bibinfo
  {journal} {IBM J. Res. Dev.}\ }\textbf {\bibinfo {volume} {1}},\ \bibinfo
  {pages} {223--231} (\bibinfo {year} {1957})}\BibitemShut {NoStop}%
\bibitem [{\citenamefont {Landauer}(1970)}]{Landauer1970_PhilMag_21_863}%
  \BibitemOpen
  \bibfield  {author} {\bibinfo {author} {\bibfnamefont {R.}~\bibnamefont
  {Landauer}},\ }\bibfield  {title} {\bibinfo {title} {{Electrical resistance
  of disordered one-dimensional lattices}},\ }\href
  {https://doi.org/10.1080/14786437008238472} {\bibfield  {journal} {\bibinfo
  {journal} {Phil Mag.}\ }\textbf {\bibinfo {volume} {21}},\ \bibinfo {pages}
  {863--867} (\bibinfo {year} {1970})}\BibitemShut {NoStop}%
\bibitem [{\citenamefont
  {B\"uttiker}(1986)}]{Buttiker1986_PhysRevLett_57_1761}%
  \BibitemOpen
  \bibfield  {author} {\bibinfo {author} {\bibfnamefont {M.}~\bibnamefont
  {B\"uttiker}},\ }\bibfield  {title} {\bibinfo {title} {{Four-Terminal
  Phase-Coherent Conductance}},\ }\href
  {https://doi.org/10.1103/PhysRevLett.57.1761} {\bibfield  {journal} {\bibinfo
   {journal} {Phys. Rev. Lett.}\ }\textbf {\bibinfo {volume} {57}},\ \bibinfo
  {pages} {1761--1764} (\bibinfo {year} {1986})}\BibitemShut {NoStop}%
\bibitem [{\citenamefont
  {B\"uttiker}(1988)}]{Buttiker1988_IBMJResDevelop_32_317}%
  \BibitemOpen
  \bibfield  {author} {\bibinfo {author} {\bibfnamefont {M.}~\bibnamefont
  {B\"uttiker}},\ }\bibfield  {title} {\bibinfo {title} {{Symmetry of
  electrical conduction}},\ }\href {https://doi.org/10.1147/rd.323.0317}
  {\bibfield  {journal} {\bibinfo  {journal} {IBM J. Res. Develop.}\ }\textbf
  {\bibinfo {volume} {32}},\ \bibinfo {pages} {317--334} (\bibinfo {year}
  {1988})}\BibitemShut {NoStop}%
\bibitem [{\citenamefont {Tanaka}\ \emph {et~al.}(2023)\citenamefont {Tanaka},
  \citenamefont {Nomoto},\ and\ \citenamefont
  {Arita}}]{Tanaka2023_PhysRevB_107_214442}%
  \BibitemOpen
  \bibfield  {author} {\bibinfo {author} {\bibfnamefont {K.}~\bibnamefont
  {Tanaka}}, \bibinfo {author} {\bibfnamefont {T.}~\bibnamefont {Nomoto}},\
  and\ \bibinfo {author} {\bibfnamefont {R.}~\bibnamefont {Arita}},\ }\bibfield
   {title} {\bibinfo {title} {{Local density of states as a probe for tunneling
  magnetoresistance effect: Application to ferrimagnetic tunnel junctions}},\
  }\href {https://doi.org/10.1103/PhysRevB.107.214442} {\bibfield  {journal}
  {\bibinfo  {journal} {Phys. Rev. B}\ }\textbf {\bibinfo {volume} {107}},\
  \bibinfo {pages} {214442} (\bibinfo {year} {2023})}\BibitemShut {NoStop}%
\bibitem [{\citenamefont {Momma}\ and\ \citenamefont
  {Izumi}(2011)}]{Momma2011_JApplCryst_44_1272}%
  \BibitemOpen
  \bibfield  {author} {\bibinfo {author} {\bibfnamefont {K.}~\bibnamefont
  {Momma}}\ and\ \bibinfo {author} {\bibfnamefont {F.}~\bibnamefont {Izumi}},\
  }\bibfield  {title} {\bibinfo {title} {{{\it VESTA3} for three-dimensional
  visualization of crystal, volumetric and morphology data}},\ }\href
  {https://doi.org/10.1107/S0021889811038970} {\bibfield  {journal} {\bibinfo
  {journal} {J. Appl. Cryst.}\ }\textbf {\bibinfo {volume} {44}},\ \bibinfo
  {pages} {1272--1276} (\bibinfo {year} {2011})}\BibitemShut {NoStop}%
\bibitem [{\citenamefont {Hirel}(2015)}]{Hirel2015ComputPhysCommun197_212}%
  \BibitemOpen
  \bibfield  {author} {\bibinfo {author} {\bibfnamefont {P.}~\bibnamefont
  {Hirel}},\ }\bibfield  {title} {\bibinfo {title} {{Atomsk: A tool for
  manipulating and converting atomic data files}},\ }\href
  {https://doi.org/https://doi.org/10.1016/j.cpc.2015.07.012} {\bibfield
  {journal} {\bibinfo  {journal} {Comput. Phys. Commun.}\ }\textbf {\bibinfo
  {volume} {197}},\ \bibinfo {pages} {212--219} (\bibinfo {year}
  {2015})}\BibitemShut {NoStop}%
\end{thebibliography}
